\documentclass[reqno,12pt,a4paper]{amsart}

\usepackage{eurosym}
\usepackage{amsfonts}
\usepackage{amssymb}
\usepackage{amsmath}
\usepackage{latexsym}
\usepackage{float,subfigure,color,subfigure}
\usepackage{graphicx}
\usepackage{amsthm}
\usepackage{verbatim}

\theoremstyle{plain}
\newtheorem{theorem}{\bf Theorem}[section]
\newtheorem{lemma}[theorem]{\bf Lemma}

\newcommand{\Sfrac}[2]{{\textstyle{\frac{#1}{#2}}}}

\newcommand{\R}{\mathbb{R}}

\DeclareMathOperator{\sech}{sech}

\title
[The Whitham equation with surface tension]{The Whitham Equation with Surface Tension}

\date{\today}

\author[Dinvay]{Evgueni Dinvay}
\author[Moldabayev]{Daulet Moldabayev}
\author[Dutykh]{Denys Dutykh}
\author[Kalisch]{Henrik Kalisch}

\address
{
	{\texttt{evgueni.dinvay@math.uib.no}},
	{\texttt{daulet.moldabayev@math.uib.no}},
	{\texttt{henrik.kalisch@math.uib.no}},
	Department of Mathematics, University of Bergen,
	Postbox 7800, 5020 Bergen, Norway.
}
\address
{
	{\texttt{Denys.Dutykh@univ-savoie.fr}},
	LAMA, UMR5127, CNRS - Universit\'{e} Savoie Mont Blanc, Campus Scientifique,
	73376 Le Bourget-du-Lac Cedex, France.
}

\begin{document}

\begin{abstract}
The viability of the Whitham equation as a nonlocal model for
capillary-gravity waves at the surface of an inviscid incompressible 
fluid is under study.
A nonlocal Hamiltonian system of model equations is derived 
using the Hamiltonian structure of the free surface water wave problem
and the Dirichlet-Neumann operator.
The system features gravitational and capillary effects, and when restricted
to one-way propagation, the system reduces to the capillary Whitham equation.

It is shown numerically that in various scaling regimes
the Whitham equation gives a more accurate approximation of the 
free-surface problem for the Euler system
than other models like the KdV, and Kawahara equation.
In the case of relatively strong capillarity considered here,
the KdV and Kawahara equations
outperform the Whitham equation with surface tension
only for very long waves with negative polarity.
\end{abstract}

\maketitle

\section{Introduction}
\setcounter{equation}{0}
We consider the water-wave problem for a layer of an incompressible
inviscid fluid bounded by a flat impenetrable bottom from below
and by a free surface from above.
The layer extends to infinity in the horizontal directions.
It is a matter of common knowledge that the Euler equations
with appropriate boundary conditions give a complete description
of the liquid dynamics.
However, in many cases the dynamics of the surface of solutions is
of particular interest.
To avoid heavy computations and concentrate attention only on
the free surface, several models approximating
evolution of the surface fluid displacement have been used.
These model equations describe only the surface dynamics without providing
complete solutions in the bulk of the fluid.
The present work focuses on the derivation of
a non-local water-wave model known as the Whitham equation
that is fully dispersive in the linear approximation.
In particular, we extend here the results of the work \cite{MKD}
to the case where surface tension is taken into account.
The model equation under study is written as
\begin{equation}
\label{dimWhitham}
	\eta_t + W \eta_x + \frac 32 \eta \eta_x = 0
	,
\end{equation}
where the convolution kernel of the operator
\(
	W \eta_x = w(-i \partial _x) \eta_x
	= \big( {\mathcal F}^{-1} w \big) * \eta_x
\)	
is given in terms of the Fourier transform by
\begin{equation}
\label{symbol}
	w(\xi) = \sqrt{ ( 1 + \varkappa \xi^2 ) \Sfrac{ \tanh(\xi) }{ \xi }}.
\end{equation}
Here it is assumed that the variables are suitably normalized so that
the gravitational acceleration, the undisturbed depth of the fluid
and the density are all unity.
The surface fluid tension is included here by means of the
capillarity parameter $\varkappa$ which is the inverse of the Bond number.
The convolution can be thought of as a Fourier multiplier operator,
and \eqref{symbol} represents the Fourier symbol of the operator. 
It is also convenient from the analytical point of view to
regard the Whitham operator $W = w(-i \partial _x)$
as an integral with respect to the spectral measure
of the self-adjoint operator $-i \partial _x$ in $L^2( \mathbb R )$.
Thus after linearization \eqref{dimWhitham} can be considered
as a Schr\"odinger equation with the self-adjoint operator
$-i \partial _x w(-i \partial _x)$.
Indeed, introducing operator $D = -i \partial _x$ one may
rewrite \eqref{dimWhitham} as
\[
	i \eta_t = D w(D) \eta + \frac 34 D \eta^2
	.
\]
From this point of view, for example, one may deduce straight away that for any
real valued solution $\eta \in C^1( \mathbb R, L^2( \mathbb R ) )$
of this equation the $L^2$-norm does not depend on time.

The Whitham equation was proposed by Whitham \cite{Wh1} as an alternative to the
well known Korteweg-de Vries (KdV) equation
\begin{equation}
\label{KdV}
	\eta_t + \eta_x + \frac 32 \eta \eta_x
	- \frac 12 \left( \varkappa - \frac 13 \right) \eta_{xxx} = 0
	.
\end{equation}
Provided $\varkappa < 1/3$ one may rescale $x$ and $t$ by
$\sqrt{1 - 3 \varkappa}$ and arrive at the equation
\[
	\eta_t + \eta_x + \frac 32 \eta \eta_x
	+ \frac 16 \eta_{xxx} = 0
	.
\]
Thus it is apparent that small capillary effect do not add anything new
to the KdV model.
However, when $\varkappa$ is near $1/3$ one cannot expect
that this model be applicable.
To describe surface waves in such a situation, one may 
use instead the fifth-order-model equation
\begin{equation}
\label{Kawahara}
	\eta_t + \eta_x + \frac 32 \eta \eta_x
	- \frac 12 \left( \varkappa - \frac 13 \right) \eta_{xxx}
	+ \frac{1}{360} ( 19 - 30 \varkappa - 45 \varkappa^2 ) \eta_{xxxxx}
	= 0
	.
\end{equation}
In our numerical experiments we use $\varkappa = 1/3$, so that the equation
reduces to what is known as the Kawahara equation \cite{Biswas,Chardard,Kawahara}, which
has the following form:
\[
	\eta_t + \eta_x + \frac 32 \eta \eta_x
	+ \frac{1}{90} \eta_{xxxxx}
	= 0
	.
\]

The validity of the KdV and Kawahara equations can be described in the terms
of the Stokes number
\(
	\mathcal S = \alpha \lambda^2
\)
where $\alpha = a/h_0$ and $\lambda/h_0$ represent a prominent amplitude and
a characteristic wavelength of the wave field respectively.
The KdV equation is known to be a good model for water waves if the amplitude of 
the waves is small and the wavelength is large when compared to the undisturbed depth,
and if in addition, the two non-dimensional quantities $\alpha$ and 
$1 / \lambda^2$ are of similar size which means $\mathcal S \sim 1$.
With the same requirement and $\varkappa$ near $1/3$ the Kawahara
equation gives better results for waves where capillarity is important.
An alternative model where the air density above the free surface
is taken into account was proposed in \cite{Stepanyants}.
Another alternative model to the KdV equation \eqref{KdV} known as the BBM equation
was put forward in \cite{Pe} and studied in depth in \cite{BBM}.
The corresponding model with the capillarity $\varkappa$ has the form
\begin{equation}
\label{BBM}
	\eta_t + \eta_x + \frac 32 \eta \eta_x
	+ \frac 12 \left( \varkappa - \frac 13 \right) \eta_{xxt} = 0
	.
\end{equation}
The linearized dispersion relation 
of this equation is not an exact match to the dispersion relation of the full water-wave problem,
but it is much closer than the KdV equation in the case when $\varkappa < 1/3$, 
and it might also be expected that this equation may be able to model shorter waves more
successfully than the KdV equation.
However, the domain of its applicability is $\mathcal S \sim 1$,
that coincides with the corresponding
restrictions of the KdV model \cite{BBM}.
One may also notice that \eqref{BBM} can also be scaled
to the equation without capillarity in the same way as \eqref{KdV}
providing capillarity $\varkappa$ is small.

Both KdV and Kawahara equations are generally believed to approximate
very long waves quite well,
but one notorious problem with these equations is that they do not model
accurately the dynamics of shorter waves.
Recognizing this shortcoming of the KdV equation, Whitham proposed to use
the same nonlinearity as the KdV equation, but coupled with a linear term
which mimics the linear dispersion relation of the full water-wave problem.
Thus, at least in theory, 
the Whitham equation can be expected to yield a 
description of the dynamics of shorter waves which is closer to
the governing Euler equations.
The Whitham equation \eqref{dimWhitham} has been studied from
a number of vantage points during recent years.
In particular, the existence of traveling and solitary waves
has been studied  \cite{AMP,EGW,EK1,EK2}.
Well posedness of a similar equation
was investigated in \cite{LannesBOOK}, and similar
full dispersion equations were also studied in \cite{LS}.
Moreover, it has been shown
in \cite{HJ1,HJ2,Sanford}
that periodic solutions of equation \eqref{dimWhitham}
feature modulational instability 
for short enough waves in a similar way as small-amplitude periodic wave
solutions of the water-wave problem. 
The performance of the Whitham equation in the description of surface water
waves has been investigated in \cite{BKN} in the steady case without surface tension.
However, it appears that no study of the performance of the Whitham equation
in the presence of capillarity has been done.

In the present note, we give an asymptotic derivation of the Whitham equation
as a model for surface water waves, 
giving close consideration to the influence of the surface tension.
The derivation proceeds by examining the Hamiltonian
formulation of the water-wave problem due to Zhakarov, Craig and Sulem \cite{Zh,CS}.
This approach is similar to the method of \cite{CG}.
However, our consideration is not constrained heavily
by any particular scalar regime.
Firstly, a corresponding Whitham system is derived, and then the Whitham
equation is found by restricting the system to one-way propagation.
Secondly, we derive different models from the Whitham equation
and point out the corresponding domains of their applicability.

Finally, a numerical comparison of modeling properties
of the KdV, Kawahara and Whitham equations is given
with respect to the Euler system.

\section{Euler system and its Hamiltonian}
\setcounter{equation}{0}
%
%
The surface water-wave problem 
is generally described by the Euler equations with no-flow conditions at
the bottom, and kinematic and dynamic boundary conditions at
the free surface. Assuming weak transverse effects, the unknowns
are the surface elevation $\eta(x,t)$, the horizontal and vertical 
fluid velocities $u_1(x,z,t)$ and $u_2(x,z,t)$, respectively,
and the pressure $P(x,z,t)$.
If the assumption of irrotational flow is made, then
a velocity potential $\phi(x,z,t)$ can be used.
Taking the undisturbed depth $h_0 = 1$ as a unit of distance, 
and the parameter $\sqrt{h_0/g} = 1$ as a unit of time, 
the problem may be posed on a domain
$\left\{(x,z) \in \R^2 | -1 < z < \eta(x,t) \right\}$
which extends to infinity in the positive and negative $x$-direction.
Due to the incompressibility of the fluid, the potential then satisfies
the Laplace's equation in this domain.
The fact that the fluid cannot penetrate the bottom is expressed by
a homogeneous Neumann boundary condition at the flat bottom.
Thus we have
\begin{eqnarray*}
	\phi_{xx} + \phi_{zz} = 0 & \mbox{in} &  -1 < z < \eta(x,t)
	\\
	\phi_z = 0              & \mbox{on} &  z=-1.
\end{eqnarray*}
The pressure is eliminated with help
of the Bernoulli equation, and the free-surface boundary conditions are formulated
in terms of the potential $\varphi$ and the surface displacement $\eta$ by
\[
	\left.
		\begin{array}{rc}
			\eta_t+\phi_x\eta_x-\phi_z
			& =0,
			\\
			\phi_t+\frac{1}{2} \big( \phi^2_x+\phi^2_z \big) + \eta
			- \varkappa \eta_{xx} \big( 1 + \eta_x^2 \big)^{- 3/2}
			& = 0, 
		\end{array}
	\right\}
	\mbox{on} \ z=\eta(x,t).
\]
The first equation represents the definition of the fluid velocity
with respect to the Euler coordinates.
The second one is the Bernoulli equation with the capillary term.

The total energy of the system is given by the sum of the kinetic energy,
the potential energy and the surface tension energy,
and normalized in such a way that the total energy
is zero when no wave motion is present at the surface.
Accordingly the Hamiltonian function for this problem is
\[
	H = \int _{\mathbb R} \int_0^\eta z \, dz dx +
	\int _{\mathbb R} \int_{-1}^\eta \Sfrac{1}{2} |\nabla \phi|^2 \, dz dx
	+ \varkappa \int _{\mathbb R} \Sfrac{ \eta_x^2 }{ 1 + \sqrt{1 + \eta_x^2} } dx	
	.
\]
Defining the trace of the potential at the free surface as
$\Phi(x,t) = \phi(x,\eta(x,t),t)$,
one may integrate in $z$ in the first integral and use the divergence theorem 
on the second integral in order to arrive at the formulation
\begin{equation}
\label{Hamiltonian-eta-phi}
	H  = \int_\R \Big[ \Sfrac{1}{2} \eta^2 + \Sfrac{1}{2} \Phi G(\eta) \Phi
	+ \varkappa \Sfrac{ \eta_x^2 }{ 1 + \sqrt{1 + \eta_x^2} }	
	\Big] \, dx.
\end{equation}
This is the Hamiltonian of the water wave problem with surface tension
as for instance found in \cite{Alazard},
and written in terms of the Dirichlet-Neumann operator $G(\eta)$.
As shown in \cite{Nicholls}, the Dirichlet-Neumann operator is analytic
in a certain sense, and can be expanded as a power series as
\[
	G(\eta)\Phi = \sum_{j=0}^\infty G_j(\eta) \Phi
	.
\]
In order to proceed, we need to understand the first few terms in this series.
As shown in \cite{CS} and \cite{CG}, the first two terms in this series
can be written with the help of the operator $D = - i \partial_x$ as 
\[
	G_0(\eta)= D\tanh(D)
	, \qquad 
	G_1(\eta)= D\eta D - D\tanh(D) \eta D\tanh(D)
	. 
\]
Note that it can be shown that the terms $G_j(\eta)$ for $j\ge 2$ are of quadratic or
higher-order in $\eta$, and will therefore not be needed in the following
analysis.

It will be convenient for the present purpose to formulate the Hamiltonian
in terms of the dependent variable $u = \Phi_x$.
This new variable is proportional to the velocity of the fluid
tangential to the surface.
More precisely
\(
	u = \varphi_x + \eta_x \varphi_z
	= \varphi _{\tau} \sqrt{1 + \eta_x^2}
\)
where $\varphi _{\tau}$ is exactly the tangential
velocity component to the surface.
To this end, we define the operator $\mathcal{K}$ by
\begin{equation*}
	G(\eta) = D \mathcal{K}(\eta) D
	.
\end{equation*}
As was the case with $G(\eta)$, the operator $\mathcal{K}(\eta)$ 
can also be expanded in a Taylor series
with respect to powers of $\eta$ and $\eta_x$ as
\[
	\mathcal{K}(\eta) = \sum_{j=0}^\infty \mathcal{K}_j(\eta)
	, \qquad
	\mathcal{K}_j(\eta) = D^{-1}G_j(\eta)D^{-1}
	.
\]
In particular, note that $\mathcal{K}_0 = \tanh{D} / D$
and
$\mathcal{K}_1 = \eta - \tanh{D} (\eta \tanh{D})$.
After integrating by parts the Hamiltonian can be expressed as 
\begin{equation}
\label{Hamiltonian_eta_u}
	H  = \int_\R \Big[ \Sfrac{1}{2} \eta^2 + \Sfrac{1}{2} u \mathcal K(\eta) u
	+ \varkappa \Sfrac{ \eta_x^2 }{ 1 + \sqrt{1 + \eta_x^2} }	
	\Big] \, dx
	.
\end{equation}
The following analysis has the formal character of long-wave approximation.
Consider a wave-field having a characteristic non-dimensional wavelength
$\lambda$ and a characteristic non-dimensional amplitude $\alpha$.
We also introduce the small parameter $\mu = \frac{1}{\lambda}$.
To obtain different approximations of the discussed problem
the amplitude $\alpha$ is considered as a function of wave-number $\mu$.
Its behavior at small wave-numbers defines different scaling regimes.
The long-wave approximation means the scale $\eta = O(\alpha)$,
$u = O(\alpha)$ and $D = - i \partial_x = O(\mu)$
where $\alpha = \alpha(\mu)$ depends on the small parameter $\mu$.
Now the Hamiltonian (\ref{Hamiltonian_eta_u}) may be simplified as follows
\begin{equation}
\label{Hamiltonian_expansion}
	H = H_g + H_c + O(\mu^2 \alpha^4)
\end{equation}
with the gravity term
\begin{equation}
\label{Hamiltonian0}
	H_g =
	\frac 12 \int_\R
	\Big[
		\eta^2 + u \Sfrac{\tanh D}{D} u + \eta u^2 -
		u \tanh{D} (\eta \tanh{D} u)
	\Big]
	dx
\end{equation}
and the capillary part
\begin{equation}
\label{Hamiltonian1}
	H_c
	=
	\varkappa \int_\R \Sfrac{ \eta_x^2 dx }{ 1 + \sqrt{1 + \eta_x^2} }	
	=
	\frac{\varkappa}{2} \int_\R \eta_x^2 dx 
	+ O(\mu^4 \alpha^4)
	.
\end{equation}
Before we continue with derivation of the Whitham equation we prove
the following lemma about integration by parts, which is certainly
well-known and we add it here only for completeness.
\begin{lemma}
\label{integration_by_parts}
	Let $f, g$ be real-valued square integrable functions
	on real axis $\mathbb R$.
	Regard $D = -i \partial_x$ as self-adjoint on $L^2(\mathbb R, \mathbb C)$
	and a real-valued function $\varphi$ that is measurable and
	almost everywhere finite with respect to Lebesgue measure.
	If $f, g$ lie in the domain of the operator $\varphi(D)$ then
	\[
		\int f \varphi(D) g = \int g \varphi(-D) f 
	\]
\end{lemma}
\begin{proof}
	It is given two proofs below.
	The first one is to regard 
	\(
		\varphi(D) = ( \mathcal F^{-1} \varphi ) *
	\)
	as	the operator of convolution
	in the sense of distribution theory
	\begin{multline*}
		\int f \varphi(D) g
		= 
		\int f(\xi) ( \mathcal F^{-1} \varphi ) (\xi - x) g(x) dx d\xi
		=
		\\
		= 
		\int f(\xi) ( \mathcal F^{-1} \varphi \circ (-id) )
		(x - \xi) g(x) dx d\xi
		=		
		\int g \varphi(-D) f
		.	
	\end{multline*}
	The second proof is to represent
	\(
		\varphi(D) = \int \varphi dE
	\)
	as the integral with respect to spectral measure $E$
	of the operator $D$.
	The corresponding projector of the interval $(\alpha, \beta)$
	is the convolution with the function
	\(
		e _{(\alpha, \beta)}(x) =
		\frac{1}{2 \pi ix} ( e^{i \beta x} - e^{i \alpha x} )
	\).
	So the replacement of $f$ and $g$ changes the corresponding spectral
	complex measure of intervals as follows
	\begin{multline*}
		\mu _{f,g}(\alpha, \beta)
		=
		(E(\alpha, \beta)f,g)
		=
		\int e _{(\alpha, \beta)}(x-y) f(y) g(x) dxdy
		=
		\\
		=
		\int e _{(-\beta, -\alpha)}(y-x) f(y) g(x) dxdy
		=
		\mu _{g,f}(-\beta, -\alpha)
	\end{multline*}
	which implies the statement of the lemma
	\[
		\int f \varphi(D) g
		= 
		\int \varphi (x)d\mu _{g,f}(x)
		=
		\int \varphi (-x)d\mu _{f,g}(x)
		=		
		\int g \varphi(-D) f
		.	
	\]
\end{proof}

\section{Derivation of the Whitham type evolution system}
\setcounter{equation}{0}
%
%

The water-wave problem can be rewritten as a Hamiltonian system using
the variational derivatives of $H$.
Making reference to \cite{CG, CGK} note that
the pair $( \eta, \Phi )$ represents the canonical variables
for the Hamiltonian function \eqref{Hamiltonian-eta-phi}.
However, it is more common to write the equations of motion
in the fluid dynamics of free surface in terms of $\eta$
and $u = \Phi_x$.
The transformation
\(
	( \eta, \Phi ) \mapsto ( \eta, u )
\)
is associated with the Jacobian
\[
	\frac{ \partial ( \eta, u ) }{ \partial ( \eta, \Phi )  }	
	=
	\begin{pmatrix}
		1 & 0
		\\
		0 & \partial_x
	\end{pmatrix}
	.
\]
Thus in terms of $\eta$ and $u$ the Hamiltonian equations have the form
\begin{equation}
\label{Hamilton_system_eta_u}
	\eta_t = - \partial_x \frac{\delta H}{\delta u} 
	, \qquad
	u_t = - \partial_x \frac{\delta H}{\delta \eta}
\end{equation}
that is not canonical since the associated structure map $J_{\eta,u}$ is symmetric:
\[
	J_{\eta,u}
	=
	\left(
		\frac{ \partial ( \eta, u ) }{ \partial ( \eta, \Phi )  }	
	\right)
	\begin{pmatrix}
		0 & 1
		\\
		-1 & 0
	\end{pmatrix}
	\left(
		\frac{ \partial ( \eta, u ) }{ \partial ( \eta, \Phi )  }	
	\right) ^*
	=
	\begin{pmatrix}
		0 & -\partial_x
		\\
		-\partial_x & 0
	\end{pmatrix}
	.
\]
We now derive a system of equations which is similar to the Whitham equation
\eqref{dimWhitham}, but admits bi-directional wave propagation.
The variational derivative ${\delta H_g} / {\delta u}$ is defined
by means of any real-valued square integrable function $h$ as follows
\begin{multline*}
	\int _{\mathbb R} \frac{\delta H_g}{\delta u} (x)h(x) dx
	=
	d_u H_g h
	=
	\left. \frac{d}{d\tau} \right| _{\tau = 0} H_g( u + \tau h, \eta )
	= \int _{\mathbb R} u \eta h dx
	+
	\\
	+
	\frac 12 \int_\R
	\Big[
		h \Sfrac{\tanh D}{D} u + u \Sfrac{\tanh D}{D} h -
		h \tanh{D} (\eta \tanh{D} u) - u \tanh{D} (\eta \tanh{D} h)
	\Big]
	dx
	.
\end{multline*}
Making use of integration by parts described in
Lemma~{\ref{integration_by_parts}} one obtains
\[
	\frac{\delta H_g}{\delta u}
	=
	\frac{\tanh D}{D} u + \eta u - \tanh{D} (\eta \tanh{D} u)
	=
	\frac{\tanh D}{D} u + \eta u + O(\mu^2 \alpha^2)
\]
and in the same way
\[
	\frac{\delta H_g}{\delta \eta}
	=
	\eta + \frac 12 u^2 + \frac 12 (\tanh{D} u)^2
	=
	\eta + \frac 12 u^2 + O(\mu^2 \alpha^2)
	.
\]
These variational derivatives were also obtained by Moldabayev and Kalisch
\cite{MKD}.
The capillary part $H_c$ defined by (\ref{Hamiltonian1})
gives the pressure $P$ on the surface
\[
	\frac{\delta H_c}{\delta \eta}
	=
	- \varkappa \frac{ \eta_{xx} }{ ( 1 + \eta_x^2 )^{\frac 32} }	
	=
	- \varkappa \eta_{xx} + O(\mu^4 \alpha^3)
	.
\]
At last the Hamilton system (\ref{Hamilton_system_eta_u})
is simplified to the Whitham system
\begin{align}
\label{sys1}
	\eta_t &=
	- \frac{\tanh D}{D} u_x - (\eta u)_x +
	\tanh{D} (\eta \tanh{D} u)_x + O(\mu^3 \alpha^3)
	, \\
\label{sys2}
	u_t &=
	- \eta_x - u u_x - (\tanh{D} u) \tanh{D} u_x
	+ \varkappa \eta_{xxx} + O(\mu^3 \alpha^3)
\end{align}
which is in line with the system obtained in \cite{MKD}
\begin{align*}
	\eta_t &=
	- \frac{\tanh D}{D} u_x - (\eta u)_x + O(\mu^3 \alpha^2)
	, \\
	u_t &=
	- \eta_x - u u_x + \varkappa \eta_{xxx} + O(\mu^3 \alpha^2)
	.
\end{align*}
%
%
%
\section{Derivation of Whitham type evolution equations}
\setcounter{equation}{0}
%
It turns out that the Whitham system \eqref{sys1}, \eqref{sys2} might be
rewritten as a system of two independent equations by further simplification.
More precisely, they will be independent with respect to
the linear approximation of that system.
For this purpose we need to separate solutions corresponding
to waves moving in other directions.
In order to derive the Whitham equation for uni-directional wave propagation,
it is important to understand how one-way propagation works in the 
Whitham system \eqref{sys1}, \eqref{sys2}.
Regard the linearisation of this system
\begin{align}
\label{linear_sys1}
	\eta_t + \frac{\tanh D}{D} u_x &= 0
	, \\
\label{linear_sys2}
	u_t + ( 1 + \varkappa D^2 ) \eta_x &= 0
	.
\end{align}
Regarding solutions of this linear system in the wave form
\[
	\eta(x, t) = A e^{ i \xi x - i \omega t }
	, \qquad
	u(x, t) = B e^{ i \xi x - i \omega t }
\]
gives rise to the matrix equation
\[
	\begin{pmatrix}
		- \omega					&	\tanh \xi
		\cr
		\xi + \varkappa \xi^3	&	- \omega	
	\end{pmatrix}
	\begin{pmatrix}
		A
		\cr
		B
	\end{pmatrix}
	=
	\begin{pmatrix}
		0
		\cr
		0
	\end{pmatrix}
	.
\]
This equation has a non-trivial solution, provided
its determinant equals zero, so that
\(
	\omega^2 - ( \xi + \varkappa \xi^3 ) \tanh \xi = 0
\).
Defining the phase speed as $c = \omega(\xi) / \xi$ one obtains
the dispersion relation 
\[
	c^2(\xi) = ( 1 + \varkappa \xi^2 ) \frac{\tanh \xi}{\xi}
\]
which coincides, up to the sign of $c$, with Whitham dispersion
relation (\ref{symbol}).
Obviously, the choice $c > 0$ corresponds to right-going wave solutions
of the linear system \eqref{linear_sys1}, \eqref{linear_sys2}.
And the phase speed $c < 0$ gives left-going waves.
To split up these two kinds of waves we regard the following
transformation of variables
\begin{equation}
\label{variable_transformation}
	r = \frac 12 (\eta + K  u)
	, \qquad
	s = \frac 12 (\eta - K  u)
	. 
\end{equation}
It is supposed that $K$ is an invertible operator,
namely an invertible function of the differential operator $D$.
The inverse transformation has the form
\begin{equation}
\label{inverse_variable_transformation}
	\eta = r + s
	, \qquad
	u = K^{-1} (r - s)
	. 
\end{equation}
The question arises whether it is possible to choose
such operator $K$ that $r$ and $s$ correspond to
right- and left-going waves, respectively.
After applying the transformation (\ref{inverse_variable_transformation})
to the linear system \eqref{linear_sys1}, \eqref{linear_sys2} one
arrives to the system
\begin{align*}
	r_t + \partial _x \big( A(D,K)r + B(D,K)s \big) &= 0
	, \\
	s_t - \partial _x \big( A(D,K)s + B(D,K)r \big) &= 0
\end{align*}
where operators $A$ and $B$ depend on $D$ and $K$ as follows
\[
	A = \frac 12 \left( (1 + \varkappa D^2) K + \frac{\tanh D}{D} K^{-1} \right)
	, \qquad
	B = \frac 12 \left( (1 + \varkappa D^2) K - \frac{\tanh D}{D} K^{-1} \right)
	.
\]
So to achieve independence of the obtained two equations
we need to choose the transformation $K$ in the way $B(D,K) = 0$,
so that
\begin{equation}
\label{K_transformation}
	K = \sqrt{ \frac{1}{ 1 + \varkappa D^2 } \cdot \frac{\tanh D}{D} }
\end{equation}
which leads to the two independent Whitham equations
\begin{align}
\label{right_linear_sys1}
	r_t + \partial _x Wr &= 0
	, \\
\label{left_linear_sys2}
	s_t - \partial _x Ws &= 0
\end{align}
where the Whitham operator $W = w(D) = A(D,K)$ was introduced at
the beginning of the paper by (\ref{symbol}).
If we again regard the wave solutions
\(
	r(x, t) = \exp({ i \xi x - i \omega_r t })
\)
and
\(
	s(x, t) = \exp({ i \xi x - i \omega_s t })
\)
then we conclude that the first equation (\ref{right_linear_sys1})
describes waves moving to the right with the phase
velocity $c_r = \omega_r / \xi = w(\xi)$
and the second equation (\ref{left_linear_sys2})
corresponds to the left-going waves with
$c_s = \omega_s / \xi = - w(\xi)$.

Now we regard the Hamiltonian (\ref{Hamiltonian_eta_u})
as a functional of $r$ and $s$
with the same long-wave approximation as before (\ref{Hamiltonian_expansion})
where obviously $r = O(\alpha)$ and
$s = O(\alpha)$.
The unperturbed Hamiltonian part (\ref{Hamiltonian0}) is
\begin{multline}
\label{Hamiltonian0_r_s}
	H_g =
	\frac 12 \int_\R
	\Big[
		(r + s)^2 + \big( K^{-1} (r - s) \big) \Sfrac{\tanh D}{D}
		K^{-1} (r - s) + (r + s) \big( K^{-1} (r - s) \big) ^2 -
	\\
		\big( K^{-1} (r - s) \big) \tanh{D} \big( (r + s)
		\tanh{D} K^{-1} (r - s) \big)
	\Big]
	dx
\end{multline}
and the surface tension adding (\ref{Hamiltonian1}) is
\begin{equation}
\label{Hamiltonian1_r_s}
	H_c
	=
	\frac{\varkappa}{2} \int_\R (r + s)_x^2 dx 
	+ O(\mu^4 \alpha^4)
	.
\end{equation}
According to the transformation theory detailed in \cite{CGK},
due to the changing of variables (\ref{variable_transformation})
or (\ref{inverse_variable_transformation}), the structure
map changes to
\[
	J_{r,s}
	=
	\left( \frac{\partial (r,s)}{\partial (\eta, u)} \right) J_{\eta, u}  
	\left( \frac{\partial (r,s)}{\partial (\eta, u)} \right)^*
	= 
	\begin{pmatrix}
		-\frac 12 \partial_x K & 0
		\\
		0 & \frac 12 \partial_x K
	\end{pmatrix}
	.
\]
The corresponding Hamiltonian system has the form
\begin{equation}
\label{Hamilton_system_r_s}
	r_t + \partial_x \left( \frac{K}{2} \frac{\delta H}{\delta r} \right) = 0
	, 
	\quad \quad 
	s_t - \partial_x \left( \frac{K}{2} \frac{\delta H}{\delta s} \right) = 0
	.
\end{equation}
These equations are equivalent to (\ref{Hamilton_system_eta_u}).
However, solutions $r(x,t)$ and $s(x,t)$ are the displacements
going right and left, respectively,
when a solution $u(x,t)$, representing the tangential velocity component
up to the curvature multiplier,
might not be imagined so easy.
As above we calculate the  G\^ateaux derivative of $H_g$
given by (\ref{Hamiltonian0_r_s})
with respect to $r$ at a real-valued square integrable function
\begin{multline*}
	\int _{\mathbb R} \frac{\delta H_g}{\delta r} (x)h(x) dx
	=
	d_r H_g h
	=
	\left. \frac{d}{d\tau} \right| _{\tau = 0} H_g( r + \tau h, s )
	=
	\int (r + s)h
	+
	\\
	+
	\frac 12 \int \big( K^{-1} h \big) \frac{\tanh D}{D} K^{-1} (r - s)
	+
	\frac 12 \int \big( K^{-1} (r - s) \big) \frac{\tanh D}{D} K^{-1} h
	+	
	\\
	+
	\frac 12 \int h	\big( K^{-1} (r - s) \big) ^2
	+
	\int (r + s) \big( K^{-1} (r - s) \big) K^{-1} h
	-
	\\
	-
	\frac 12 \int \big( K^{-1} h \big)
	\tanh{D} \big( (r + s) \tanh{D} K^{-1} (r - s) \big)
	-
	\\
	-
	\frac 12 \int \big( K^{-1} (r - s) \big)
	\tanh{D} \big( h \tanh{D} K^{-1} (r - s) \big)
	-
	\\
	-
	\frac 12 \int \big( K^{-1} (r - s) \big)
	\tanh{D} \big( (r + s) \tanh{D} K^{-1} h \big)
	.
\end{multline*}
Integrating by parts as in Lemma~{\ref{integration_by_parts}}
and taking into account that functions $\tanh D$ is odd while
$K$ is even (\ref{K_transformation}) with respect to $D$
one can obtain ${\delta H_g}/{\delta r}$,
and in the similar way ${\delta H_c}/{\delta r}$, ${\delta H_g}/{\delta s}$,
${\delta H_c}/{\delta s}$.
Thus as a result
\begin{multline}
\label{right_variational_derivative}
	\frac{K}{2} \frac{\delta H}{\delta r}
	=
	Wr + \frac 14 K \big( K^{-1} (r - s) \big)^2 +
	\frac 12 (r + s) K^{-1} (r - s)
	+
	\\
	+
	\frac 14 K \big( \tanh D K^{-1} (r - s) \big)^2 -
	\frac 12 \tanh D \big( (r + s) \tanh D K^{-1} (r - s) \big)
	+ O(\mu^2 \alpha^3)
\end{multline}
and
\begin{multline}
\label{left_variational_derivative}
	\frac{K}{2} \frac{\delta H}{\delta s}
	=
	Ws + \frac 14 K \big( K^{-1} (r - s) \big)^2 -
	\frac 12 (r + s) K^{-1} (r - s)
	+
	\\
	+
	\frac 14 K \big( \tanh D K^{-1} (r - s) \big)^2 +
	\frac 12 \tanh D \big( (r + s) \tanh D K^{-1} (r - s) \big)
	+ O(\mu^2 \alpha^3)
\end{multline}
together with (\ref{Hamilton_system_r_s}) are the Whitham system
describing the displacement $\eta(x, t)$ in terms of waves going
right and left.
This system is entirely equivalent to (\ref{sys1}), (\ref{sys2})
and at the same time gives rise to solutions with more clear physical meaning.
Another useful property of this system is that it can be enough to
regard only one equation if we are allowed to neglect the waves
going to a particular direction.
More precisely, regarding only right-going waves leads to
\begin{multline}
\label{only_right_variational_derivative}
	\frac{K}{2} \frac{\delta H}{\delta r}
	=
	Wr + \frac 14 K \big( K^{-1} r \big)^2 +
	\frac 12 r K^{-1} r
	+
	\\
	+
	\frac 14 K \big( \tanh D K^{-1} r \big)^2 -
	\frac 12 \tanh D \big( r \tanh D K^{-1} r \big)
	+ O(\alpha |s|) + O(\mu^2 \alpha^3)
\end{multline}
and only left-going waves gives
\begin{multline}
\label{only_left_variational_derivative}
	\frac{K}{2} \frac{\delta H}{\delta s}
	=
	Ws + \frac 14 K \big( K^{-1} s \big)^2 +
	\frac 12 s K^{-1} s
	+
	\\
	+
	\frac 14 K \big( \tanh D K^{-1} s \big)^2 -
	\frac 12 \tanh D \big( s \tanh D K^{-1} s \big)
	+ O(\alpha |r|) + O(\mu^2 \alpha^3)
	.
\end{multline}
As one can see these expressions are identical up to discarded parts.
Equality (\ref{only_right_variational_derivative}) together with the
first equation of (\ref{Hamilton_system_r_s}) corresponds to the
Whitham equation describing right-going surface waves.
Equality (\ref{only_left_variational_derivative}) together with the
second equation of (\ref{Hamilton_system_r_s}) corresponds to the
Whitham equation describing left-going surface waves.
Further simplifications can be made by studying
concrete regimes $\alpha(\mu)$.
As a matter of fact, examples of behavior $\alpha(\mu)$ at small $\mu$
that we regard below need less accurate asymptotic.
So that operators $K$ and $\tanh D$ can still be simplified by
taking into account $D = O(\mu)$ in Equality
(\ref{only_right_variational_derivative}) as follows
\begin{multline}
\label{simplified_variational_derivative}
	\frac{K}{2} \frac{\delta H}{\delta r}
	=
	Wr + \frac 34 r^2 +
	\frac 14 \left( \varkappa - \frac 53 \right) r D^2 r
	-
	\frac 14 \left( \varkappa + \frac 43 \right) (D r)^2
	+
	\\
	+
	O(\mu^4 \alpha^2) + O(\alpha |s|) + O(\mu^2 \alpha^3)
	.
\end{multline}
It is in line with \cite{MKD} for liquids
without surface tension $\varkappa = 0$.
There is the same expression for the other variational derivative
with replacement $r$ by $s$.
\subsection{Linear approximation.}
Let $r$ and $s$ be of the same order and $\alpha = o(1)$ as $\mu \to 0$.
Then (\ref{only_right_variational_derivative}),
(\ref{only_left_variational_derivative}) are simplified to
\[
	\frac{K}{2} \frac{\delta H}{\delta r} = Wr (1 + o(1))
	, \qquad
	\frac{K}{2} \frac{\delta H}{\delta s} = Ws (1 + o(1))
\]
that together with (\ref{Hamilton_system_r_s}) represent two
independent linear equations.
We again arrived to (\ref{right_linear_sys1}), (\ref{left_linear_sys2}).
This is the case when we do not have enough information about
order relation between right- and left-going waves.
\subsection{The shallow-water scaling regime.}
Let $\alpha = O(1)$ as $\mu \to 0$.
Assume also that left-going waves can be discarded $s = o(1)$.
In this case operators $W$ can also be simplified by
taking into account $D = O(\mu)$.
Expression (\ref{simplified_variational_derivative}) becomes
\[
	\frac{K}{2} \frac{\delta H}{\delta r}
	= r + \frac 34 r^2 + o(1)
\]
which leads to the shallow-water equation
\[
	r_t + r_x + \frac 32 r r_x = o(\mu)
	.
\]
\subsection{The Boussinesq scaling regime.}
Let $\alpha = O(\mu^2)$ and $s = o(\mu^2)$ as $\mu \to 0$.
Expression (\ref{simplified_variational_derivative}) becomes
\[
	\frac{K}{2} \frac{\delta H}{\delta r}
	= Wr + \frac 34 r^2 + o(\mu^4)
\]
that can be simplified further by two different ways
\[
	W = 1 + \frac 12 \left( \varkappa - \frac 13 \right) D^2 + O(\mu^4) 
	= \left( 1 - \frac 12 \left( \varkappa - \frac 13 \right) D^2 \right)
	^{-1} + O(\mu^4)
	.
\]
The first equality leads to the KdV equation
\[
	r_t + r_x - \frac 12 \left( \varkappa - \frac 13 \right) r_{xxx}
	+ \frac 32 r r_x = o(\mu^5)
	.
\]
The second equality gives rise to the BBM equation
\[
	r_t + r_x + \frac 12 \left( \varkappa - \frac 13 \right) r_{xxt}
	+ \frac 32 r r_x = o(\mu^5)
	.
\]
\subsection{The Pad\'e (2,2) approximation.}
Suppose $\alpha = O(\mu^4)$ and $s = o(\mu^4)$ as $\mu \to 0$.
Expression (\ref{simplified_variational_derivative}) becomes
\[
	\frac{K}{2} \frac{\delta H}{\delta r}
	= Wr + \frac 34 r^2 + o(\mu^8)
\]
that can be simplified further provided $\varkappa \neq 1/3$ by the way
\[
	W = \frac{ 1 + aD^2 }{ 1 + bD^2 } + O(\mu^6) 
\]
where constants $a$ and $b$ depend on $\varkappa$ as follows
\[
	a(\varkappa) = \frac
	{ 3 + 10 \varkappa - 45 \varkappa^2 }
	{ 20 (1 - 3 \varkappa) }
	,	
\]
\[
	b(\varkappa) = \frac
	{ 19 - 30 \varkappa - 45 \varkappa^2 }
	{ 60 (1 - 3 \varkappa) }
	.	
\]
The corresponding equation
\[
	r_t + r_x - a(\varkappa) r_{xxx} - b(\varkappa) r_{xxt}
	+ \frac 32 r r_x = o(\mu^9)
	.
\]
As one can see the order of this differential equation is the same
as the order of KdV or BBM, meanwhile the Pad\'e approximation
is more accurate.
In the case $\varkappa = 1/3$ one has to use
the usual Taylor approximation 
\[
	W = 1 + \frac 12 \left( \varkappa - \frac 13 \right) D^2
	+ \frac{1}{360} ( 19 - 30 \varkappa - 45 \varkappa^2 ) D^4 + O(\mu^6) 
\]
which gives rise to the equation of fifth order
\[
	r_t + r_x - \frac 12 \left( \varkappa - \frac 13 \right) r_{xxx}
	+ \frac{1}{360} ( 19 - 30 \varkappa - 45 \varkappa^2 ) r_{xxxxx}
	+ \frac 32 r r_x = o(\mu^9)
\]
that is the Kawahara equation \eqref{Kawahara}.
\subsection{The Whitham scaling regime.}
If we now assume $\alpha = O(\mu^N)$ for any positive integer $N$
and $s = o(\alpha)$ as $\mu \to 0$, then we arrive to an example
when the Whitham operator $W$ cannot be approximated
using a simple differential operator instead.
The simplest equation in this case is the Whitham equation
\[
	r_t + W r_x + \frac 32 r r_x = o(\mu \alpha^2)
	.
\]
An example of the function $\alpha(\mu)$ when the Whitham equation
works better than its approximations was given in \cite{MKD}.
At the same time that function $\alpha(\mu)$ may be similar
to the Boussinesq scale at some wave-numbers $\mu$
as was pointed out in \cite{MKD}.
%
%
\section{Numerical results}
\setcounter{equation}{0}
%
%
%
The purpose of this section is to compare the performance of the Whitham
equation as a model for surface water waves to both the KdV equation \eqref{KdV}
and to the Kawahara equation \eqref{Kawahara}.
In other words, all these approximate models are compared to
the Euler system which is considered as giving the closest description
of an actual surface wave profile.
For this purpose initial data are imposed, the Whitham,
KdV and Kawahara equations are solved with periodic boundary conditions,
and the solutions are compared to the numerical solutions of the full Euler equations
with free-surface boundary conditions.
This matching is made in various scaling regimes
from small Stokes numbers to $\mathcal S \sim 1$, to large Stokes numbers.

\begin{table}
\begin{center}
	\begin{tabular}{| c | c | c | c |}
		\hline
		Experiment  &  Stokes number $\mathcal S$
		&
		Amplitude $\alpha$  &  Wavelength $\lambda$
		\\
		\hline
		A	&  1   &  	 0.1 &	$\sqrt{10}$
		\\	
		B	&  1   & 	 0.2 &	$\sqrt{5}$ 			
		\\
		C	&  10  & 	 0.1 &	10	
		\\
		D	&  10  & 	 0.2 &	$\sqrt{50}$	
		\\	
		E	&  0.1 & 	 0.1 &	$1$ 	
		\\
		F	&  0.1 & 	 0.2 &	$\sqrt{ 1/2 }$			
		\\
		\hline
	\end{tabular}
\end{center}
\caption{Summary of the Stokes number, nondimensional wavelength,
nondimensional amplitude of the initial data used in the numerical experiments.}
\end{table}

The numerical treatment of the three model equations is a standard spectral scheme,
such as used in \cite{FW} and \cite{EK2} for example. For the time stepping,
an efficient fourth-order implicit method developed in \cite{FruSan} is used. 
The numerical discretization of the free-surface problem for the Euler equations
is based on a conformal mapping of the fluid domain into a rectangle. 
In the case of transient dynamics, this method has roots
in the work of Ovsyannikov \cite{Ovsyannikov}, 
and was later used in \cite{DyachenkoZakharov} and \cite{LHCh}.
In the case of periodic
boundary conditions, a Fourier-spectral collocation method can be used
for the computations, and the particular method used for the numerical experiments
reported here is detailed in \cite{MDC}.

\begin{figure}[t!]
	\centering
	\subfigure
	{\includegraphics[width=0.75\textwidth]{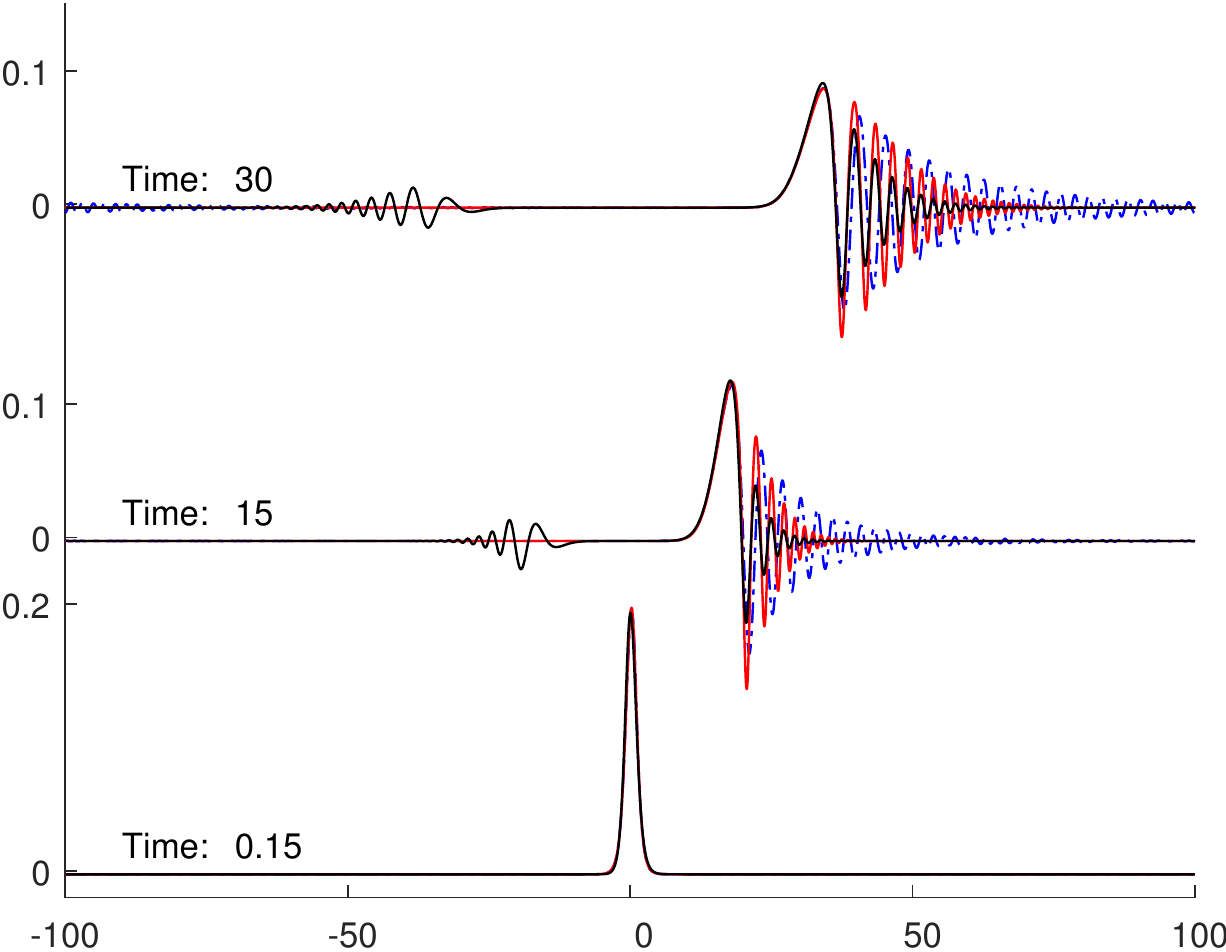}}
	\subfigure
	{\includegraphics[width=0.75\textwidth]{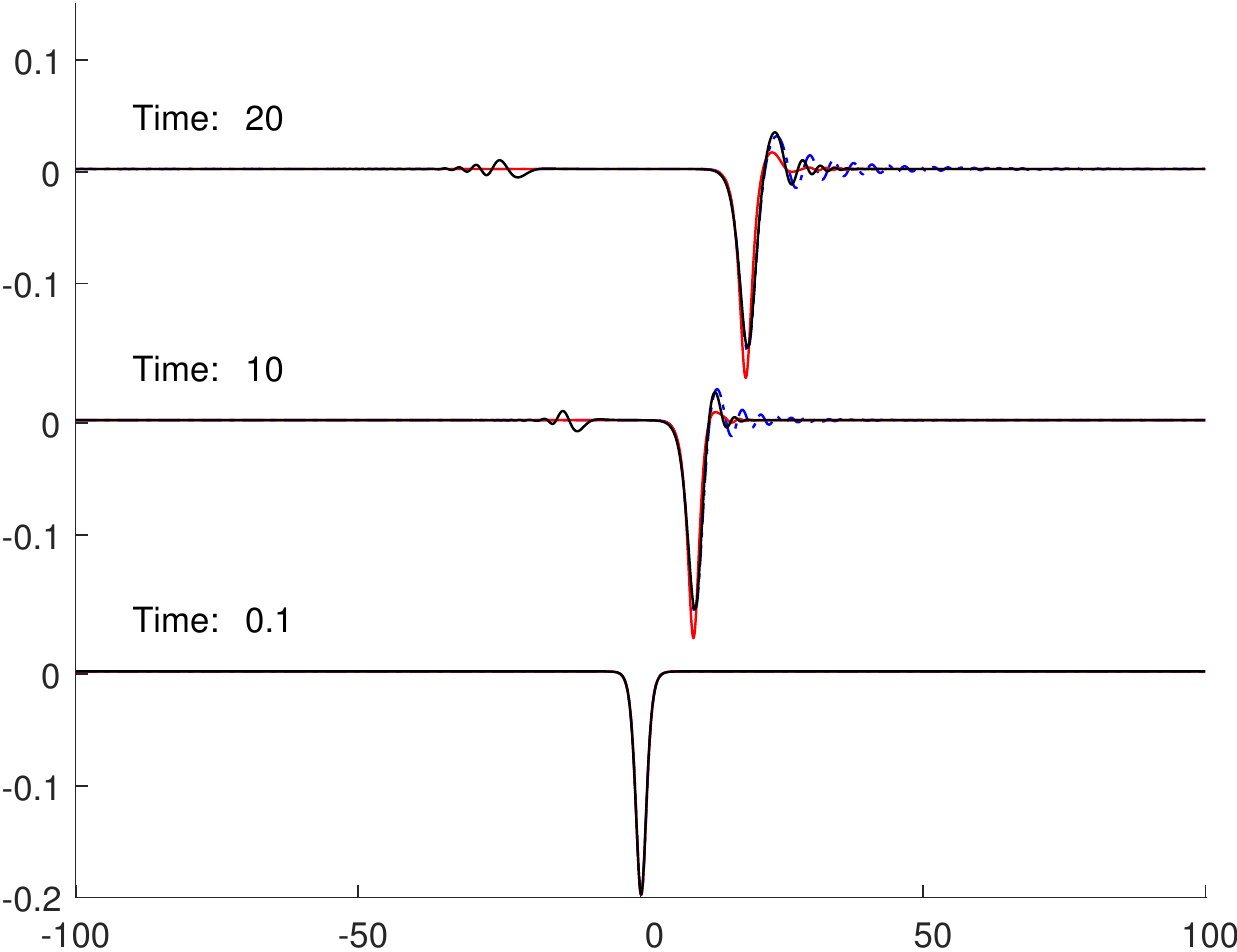}}
	\caption
	{
		Wave profiles at three different times:
		\textcolor{black}{\textbf{--}} the Euler (black line),
		\textcolor{red}{\textbf{--}} Whitham (red line)
		and \textcolor{blue}{\textbf{--}} KdV (blue line)
		with 
		amplitude $\alpha = 0.2$,
		wavelength $\lambda = \sqrt{5}$ and capillarity parameter $\varkappa = 1/2$.
	}
\label{motion_plot}
\end{figure}
\begin{figure}[t!]
	\centering
	\subfigure
	{
		\includegraphics
		[
			width=0.39\textwidth
			,
			trim = 4.0cm 9cm 4.0cm 9cm
		]
		{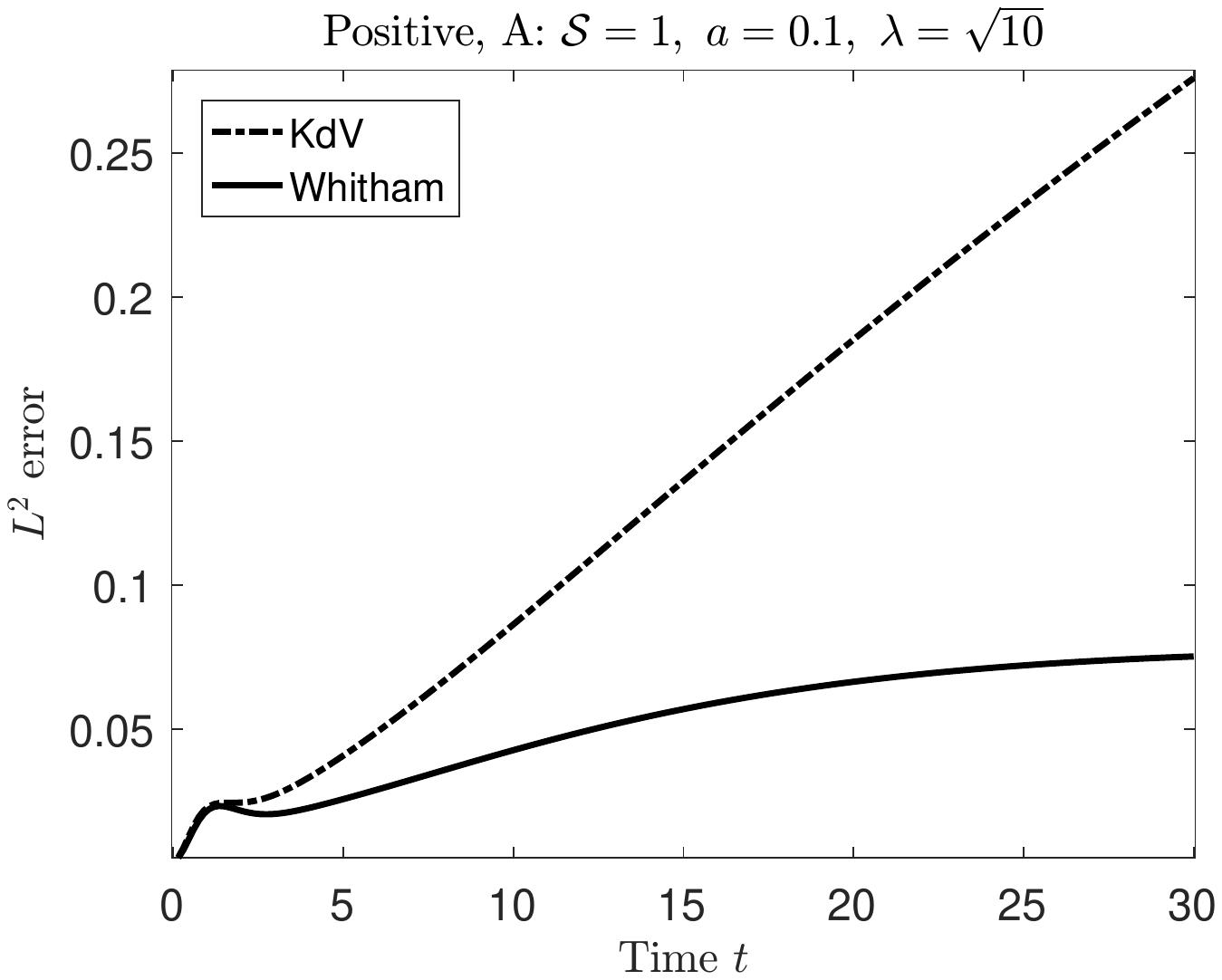}
	}
	~~~~
	\subfigure
	{
		\includegraphics
		[
			width=0.39\textwidth
			,
			trim = 4.0cm 9cm 4.0cm 9cm
		]
		{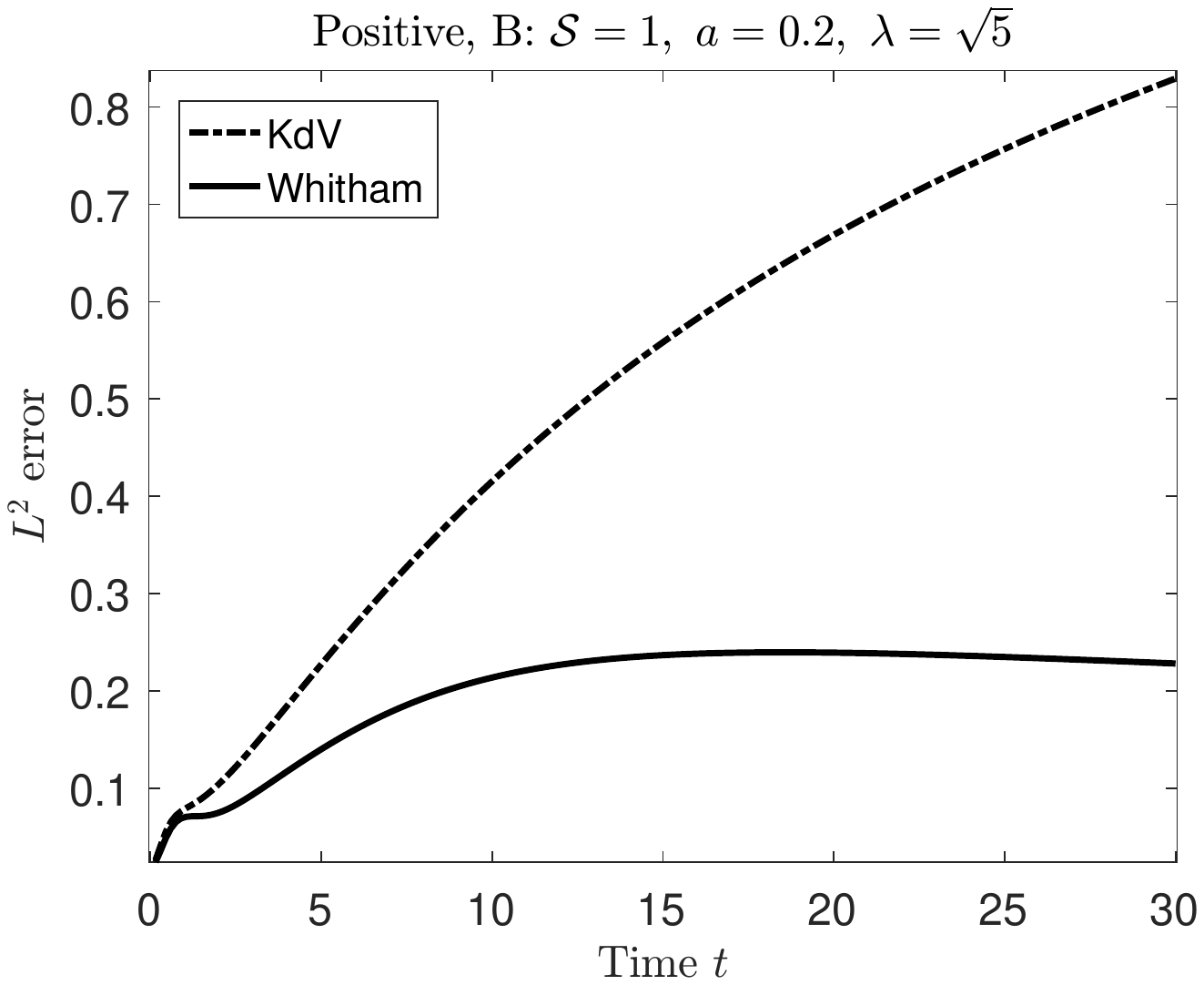}
	}
	\subfigure
	{
		\includegraphics
		[
			width=0.39\textwidth
			,
			trim = 4.0cm 9cm 4.0cm 9cm
		]
		{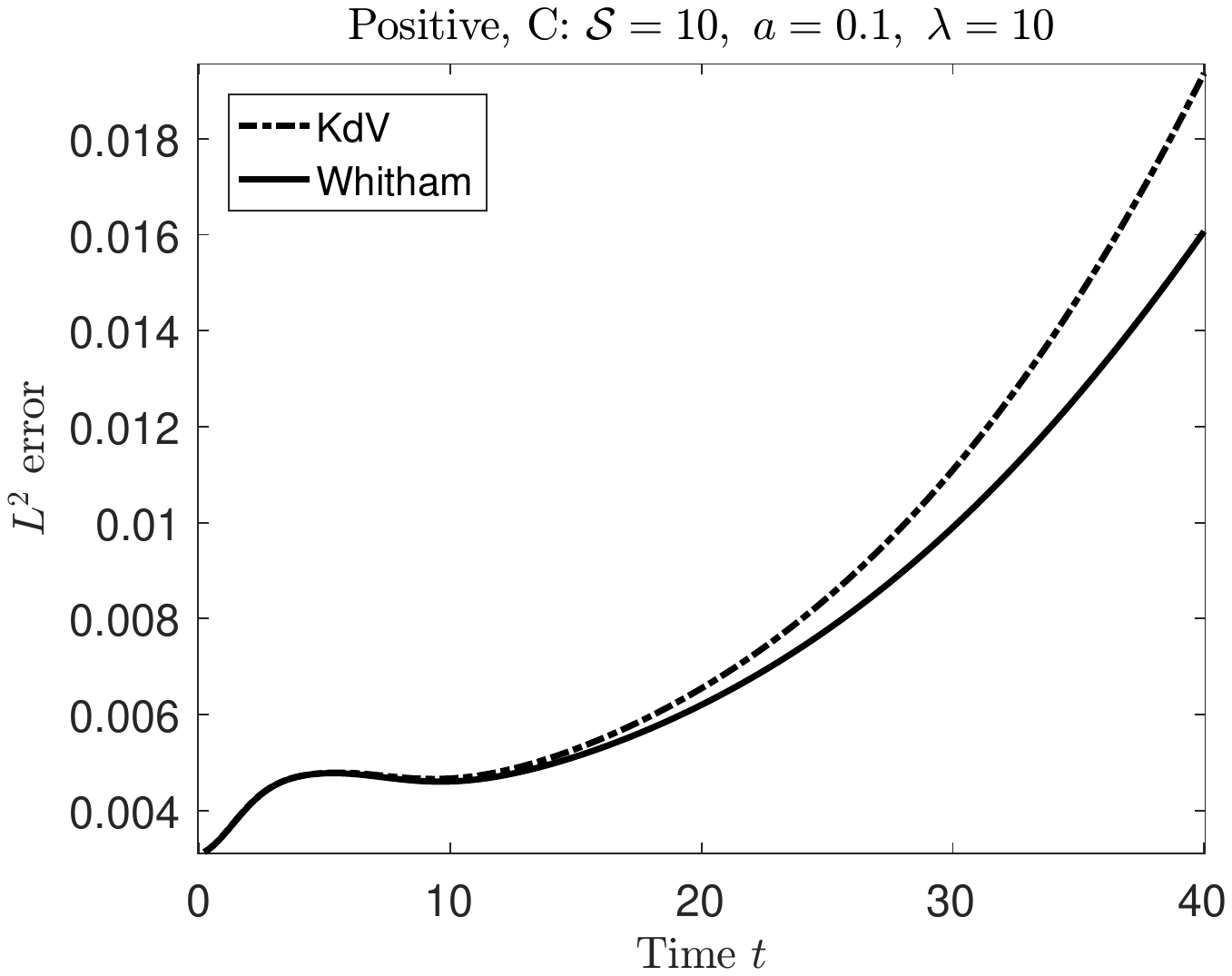}
	}
	~~~~
	\subfigure
	{
		\includegraphics
		[
			width=0.39\textwidth
			,
			trim = 4.0cm 9cm 4.0cm 9cm
		]
		{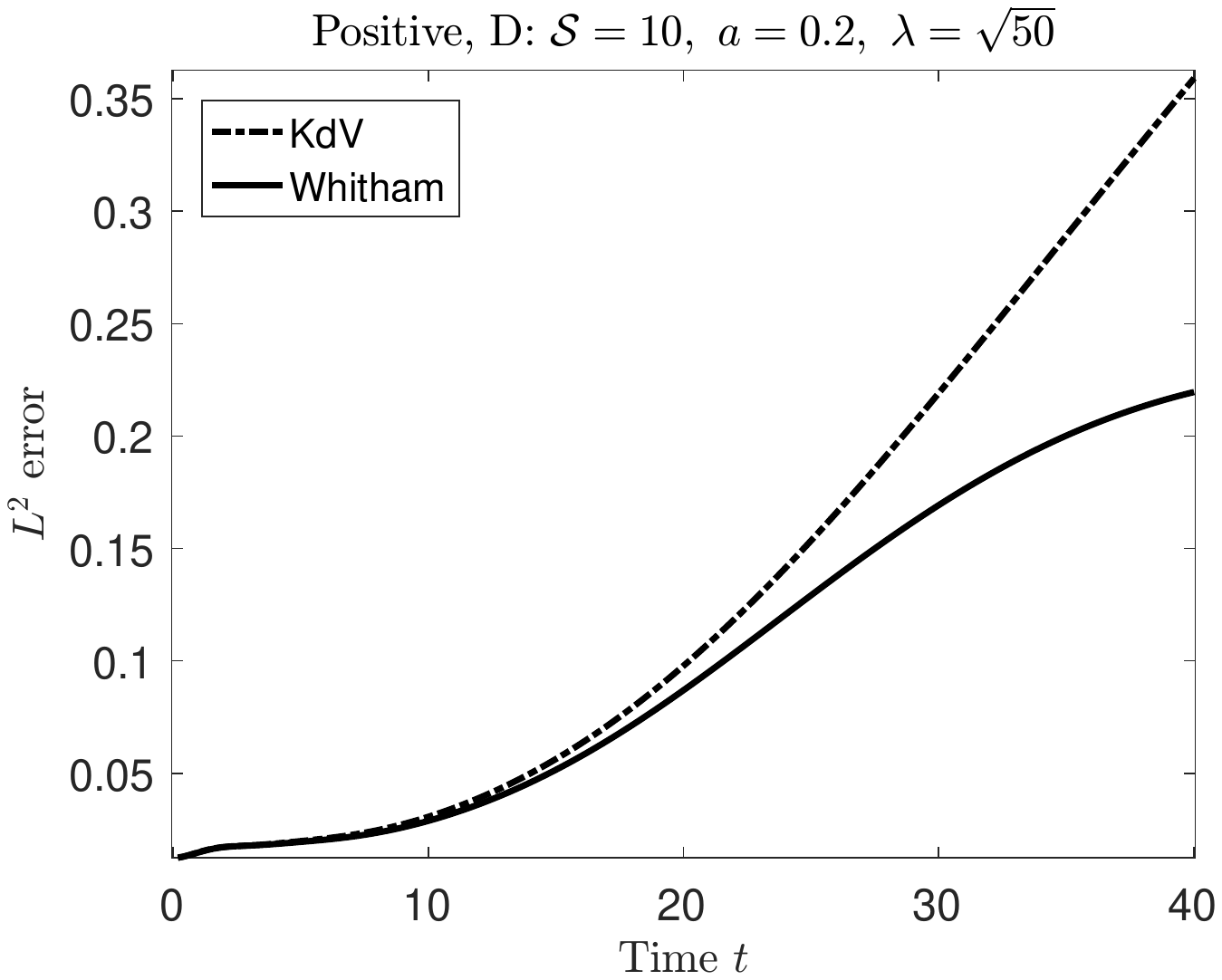}
	}
	\subfigure
	{
		\includegraphics
		[
			width=0.39\textwidth
			,
			trim = 4.0cm 9cm 4.0cm 9cm
		]
		{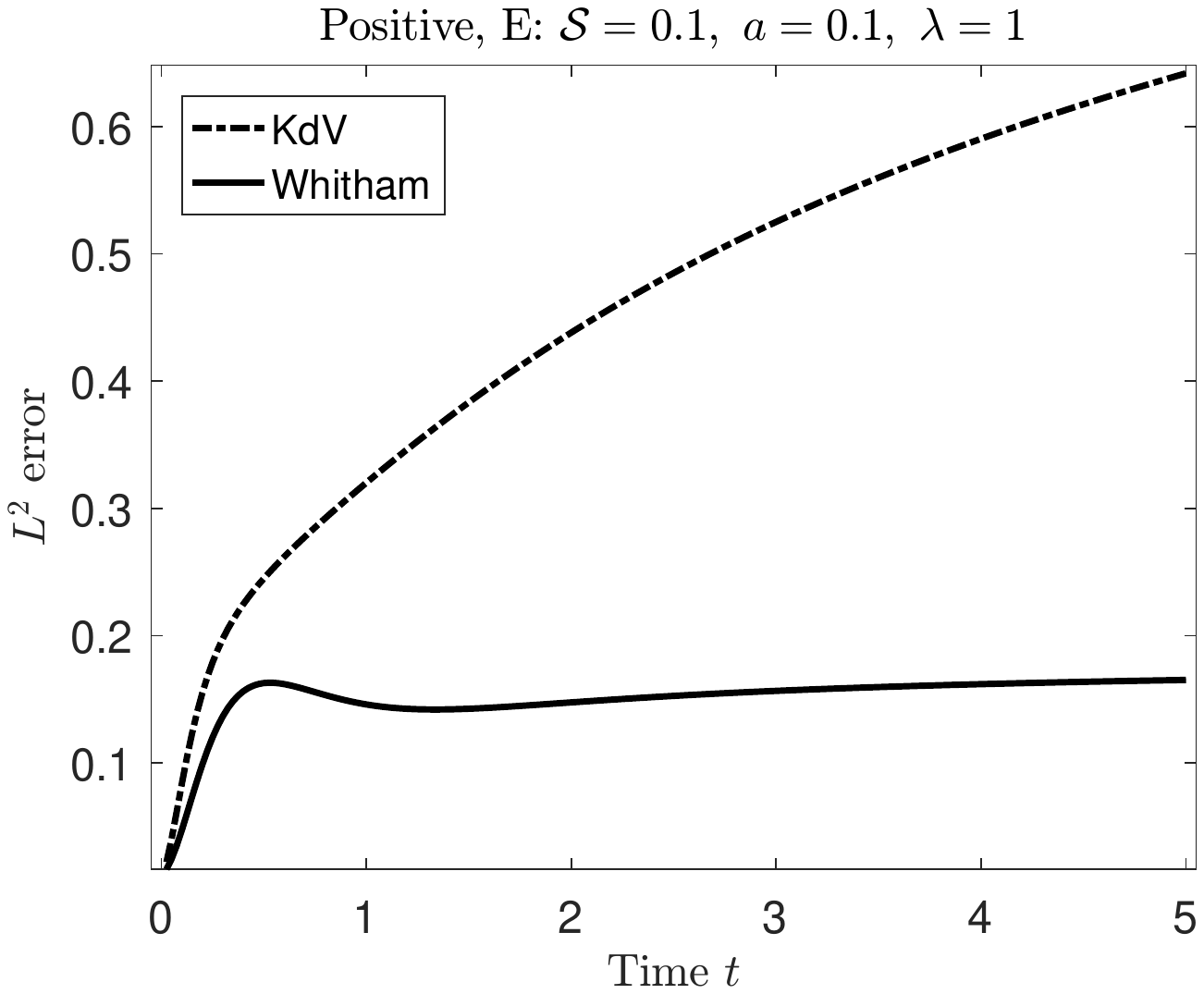}
	}
	~~~~
	\subfigure
	{
		\includegraphics
		[
			width=0.39\textwidth
			,
			trim = 4.0cm 9cm 4.0cm 9cm
		]
		{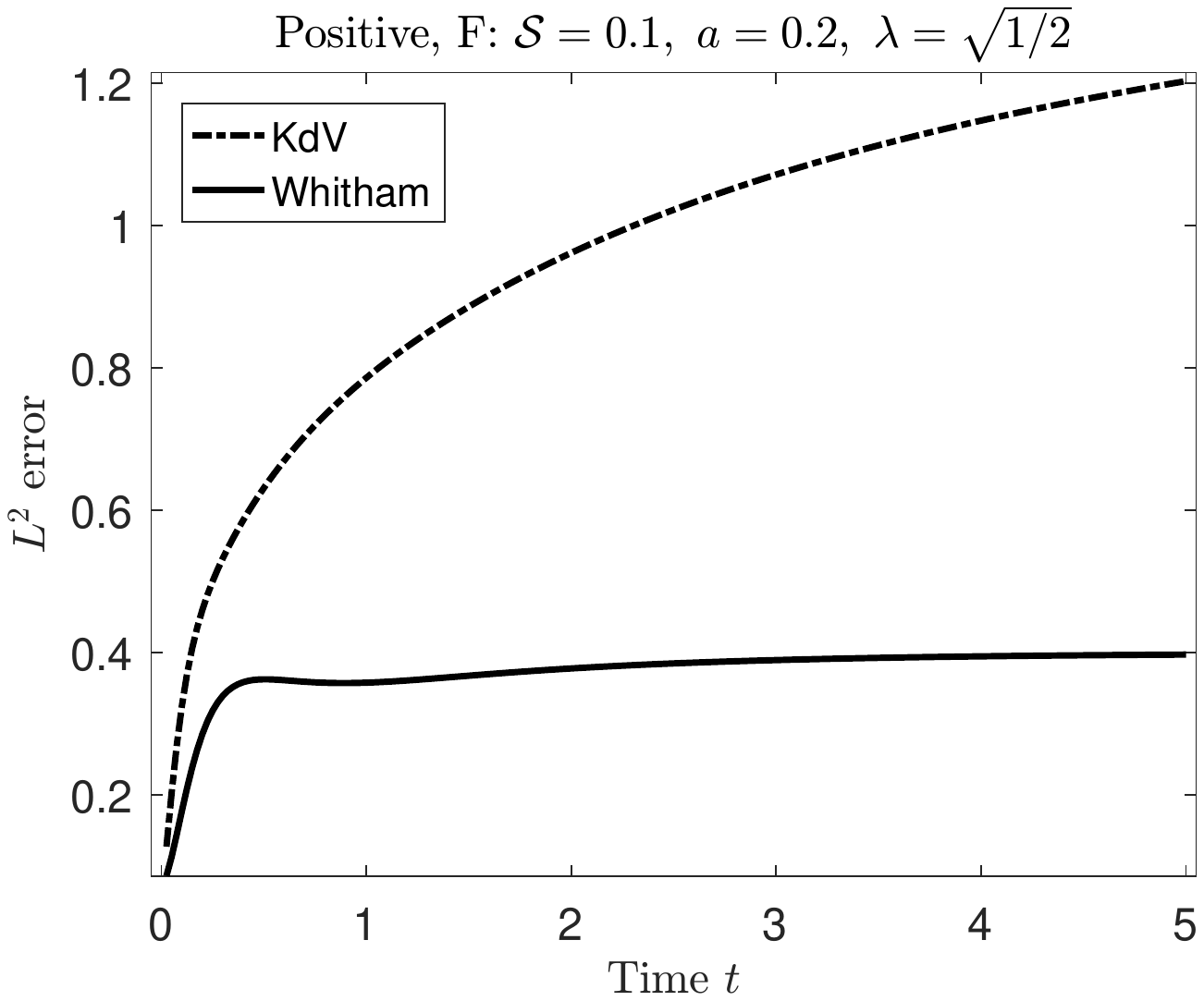}
	}
	\caption
	{
		$L^2$ errors in approximation of solutions
		to the full Euler equations by different model equations
		with the positive initial wave $\eta_0(x)$ and
		the surface tension $\varkappa = \frac 12$.
	}
\label{error_positive_plot_with_tension_frac_12}
\end{figure}
\begin{figure}[t!]
	\centering
	\subfigure
	{
		\includegraphics
		[
			width=0.39\textwidth
			,
			trim = 4.0cm 9cm 4.0cm 9cm
		]
		{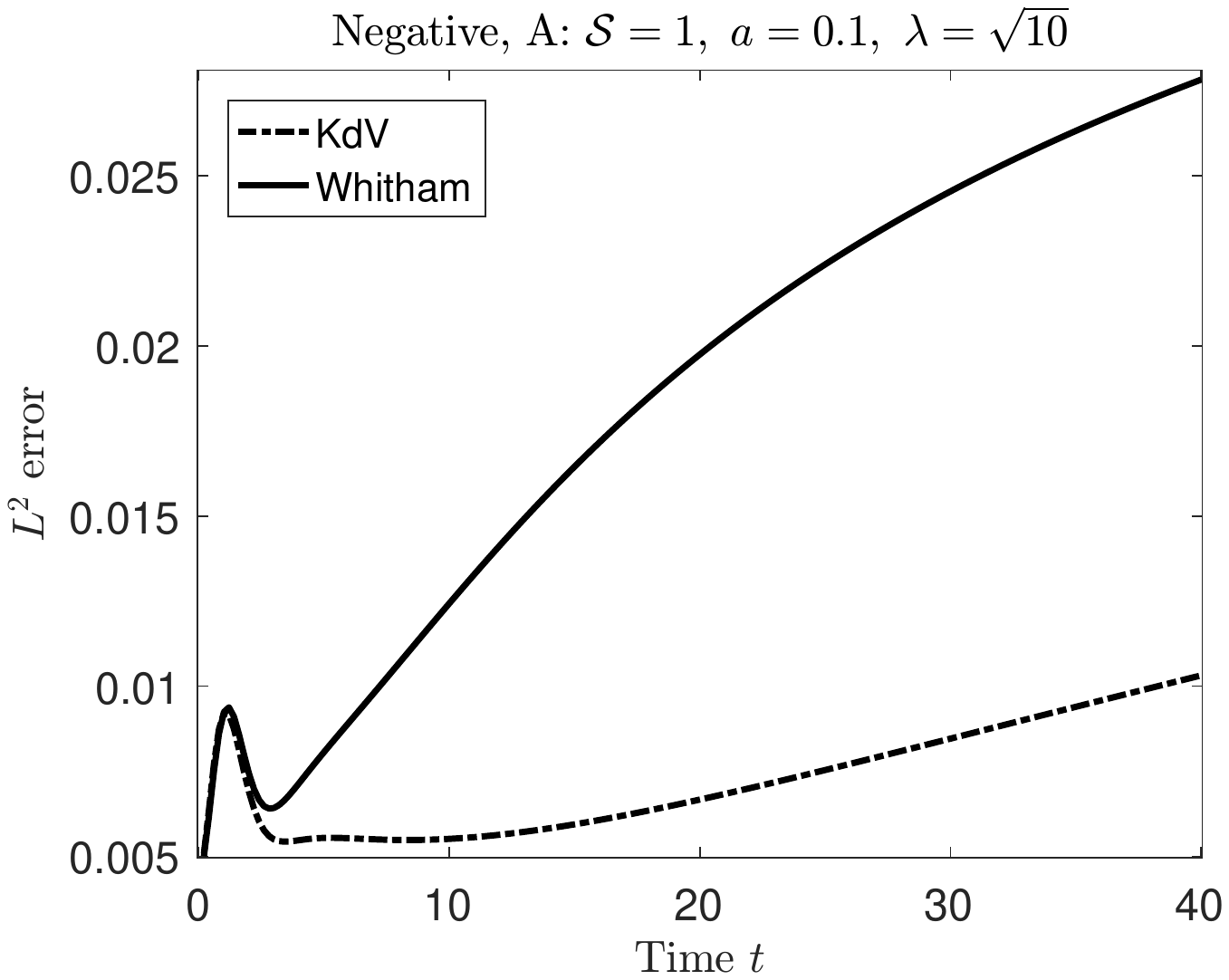}
	}
	~~~~
	\subfigure
	{
		\includegraphics
		[
			width=0.39\textwidth
			,
			trim = 4.0cm 9cm 4.0cm 9cm
		]
		{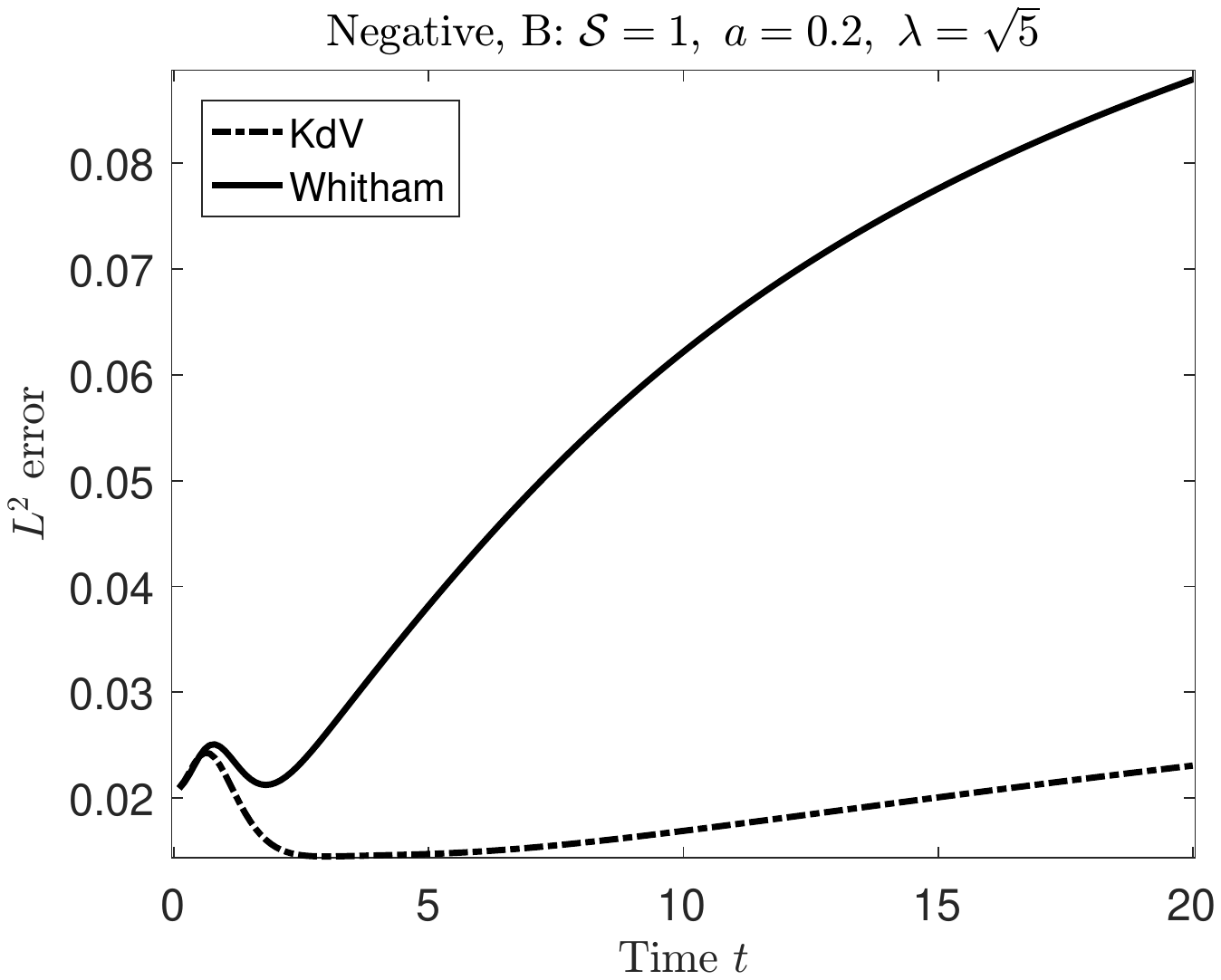}
	}
	\subfigure
	{
		\includegraphics
		[
			width=0.39\textwidth
			,
			trim = 4.0cm 9cm 4.0cm 9cm
		]
		{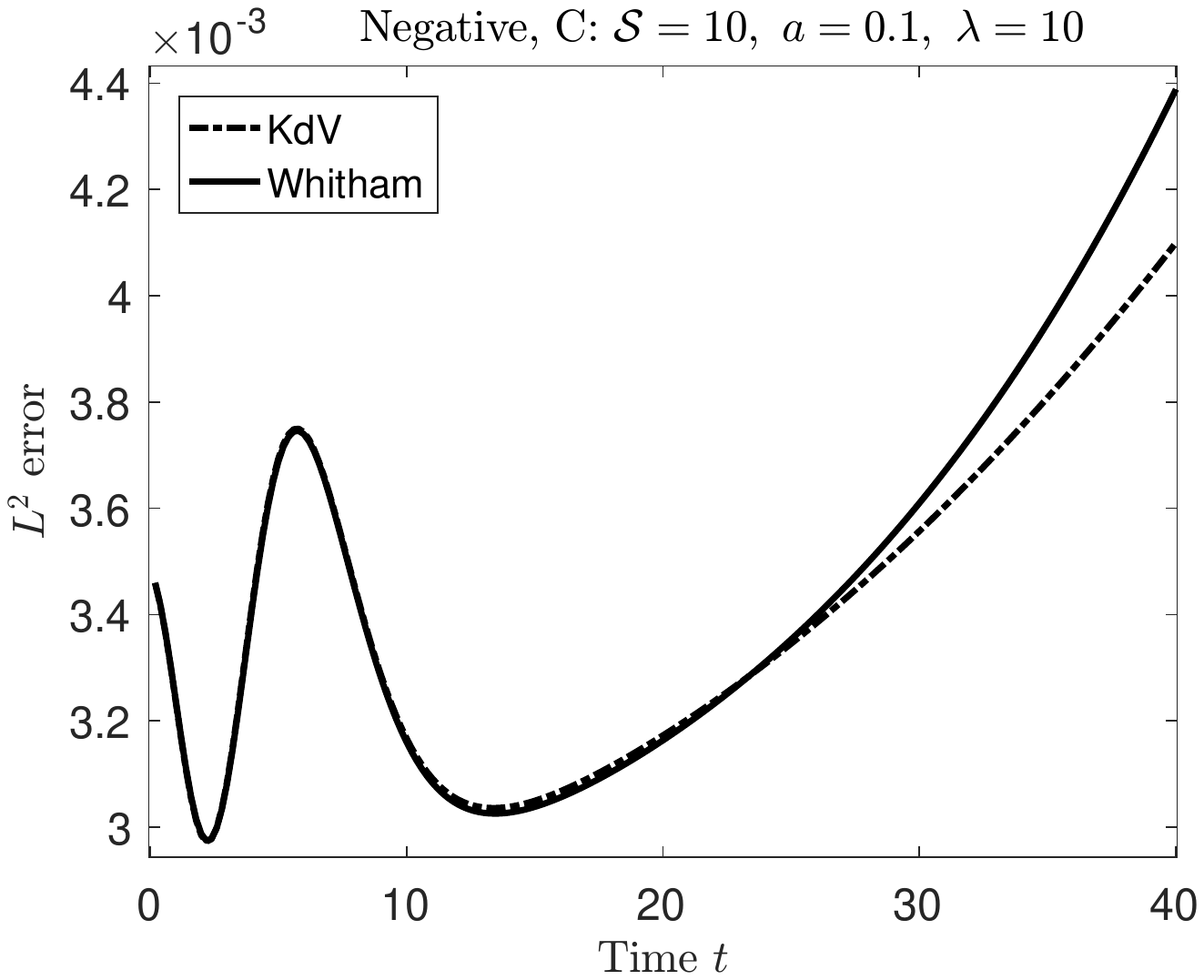}
	}
	~~~~
	\subfigure
	{
		\includegraphics
		[
			width=0.39\textwidth
			,
			trim = 4.0cm 9cm 4.0cm 9cm
		]
		{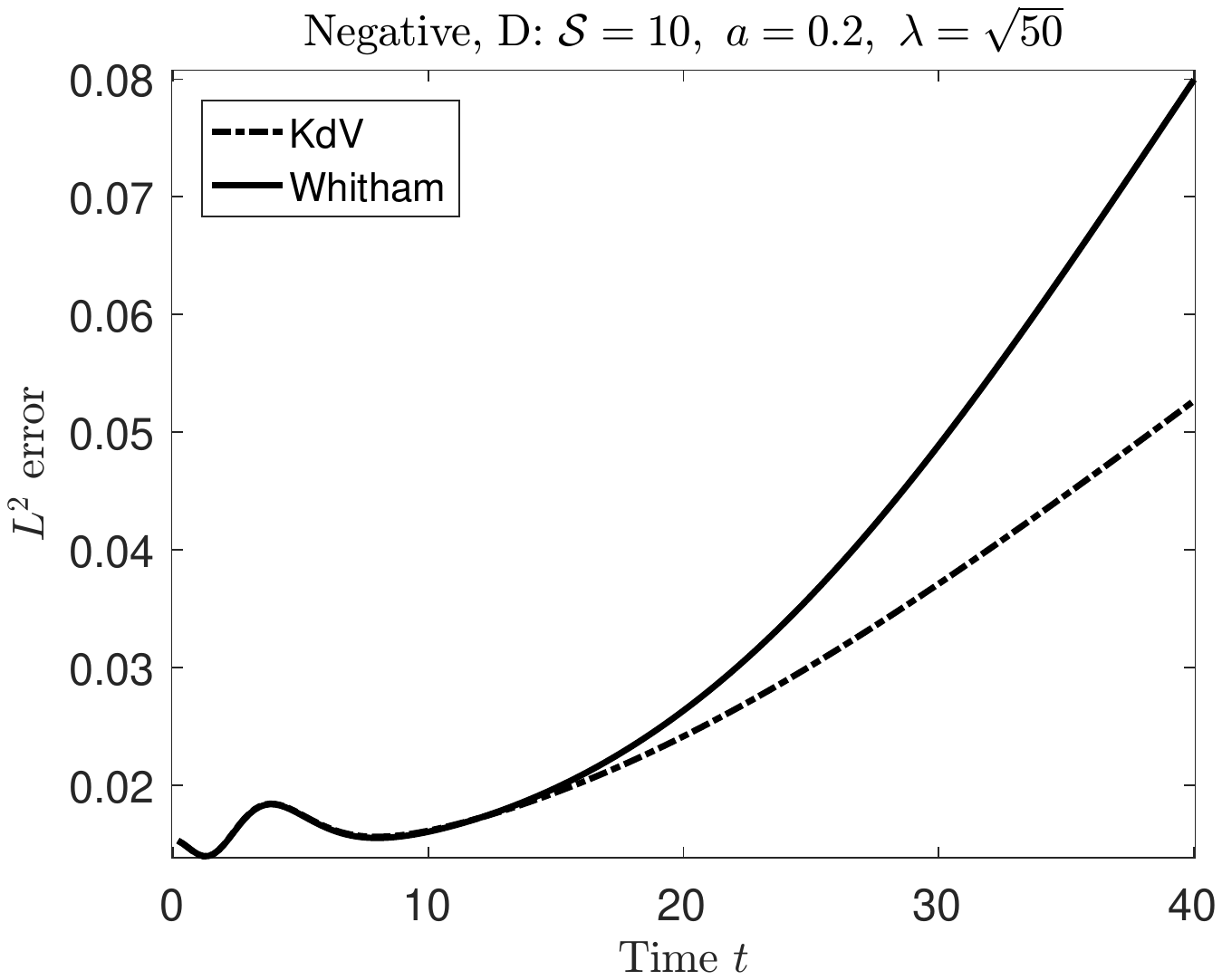}
	}
	\subfigure
	{
		\includegraphics
		[
			width=0.39\textwidth
			,
			trim = 4.0cm 9cm 4.0cm 9cm
		]
		{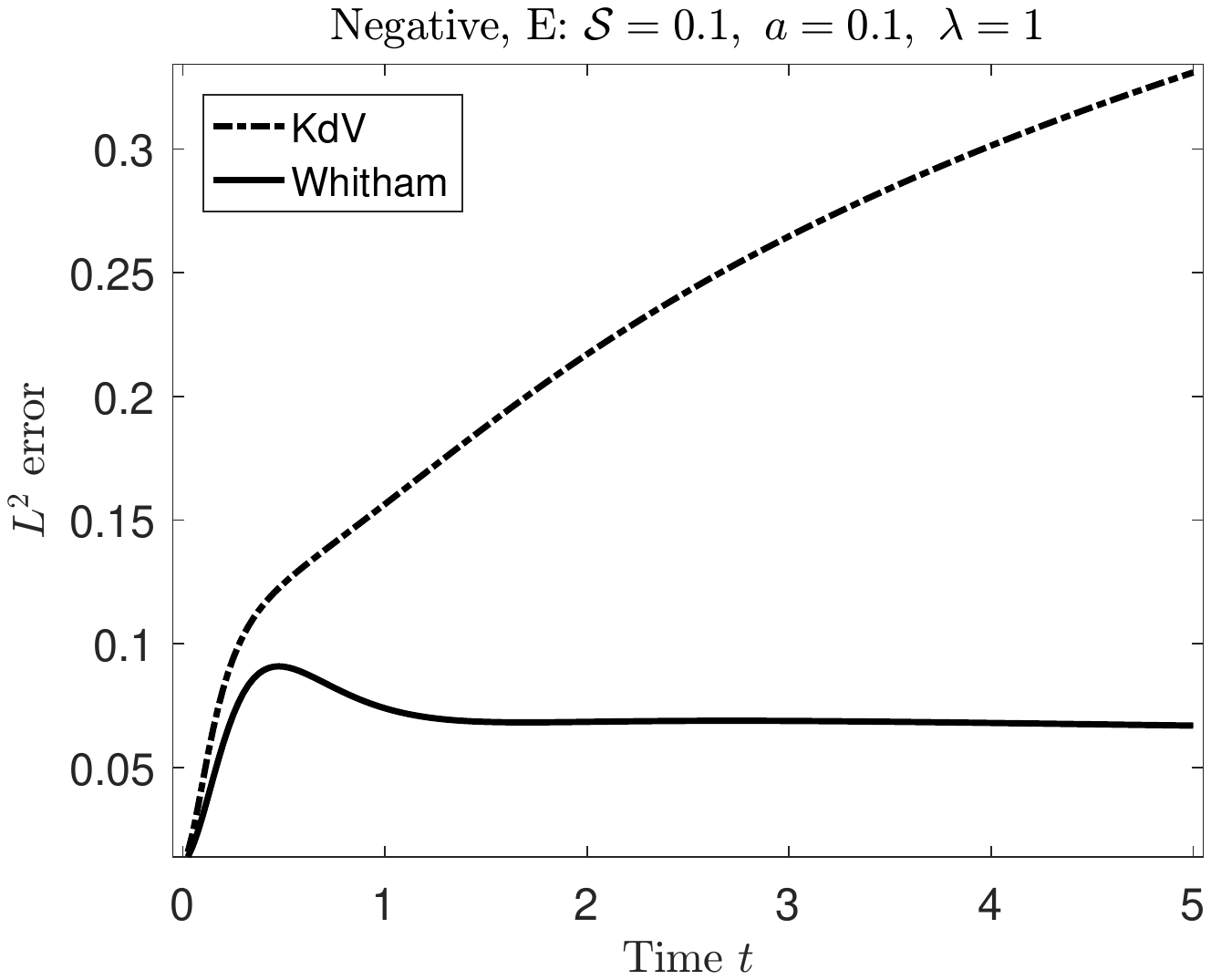}
	}
	~~~~
	\subfigure
	{
		\includegraphics
		[
			width=0.39\textwidth
			,
			trim = 4.0cm 9cm 4.0cm 9cm
		]
		{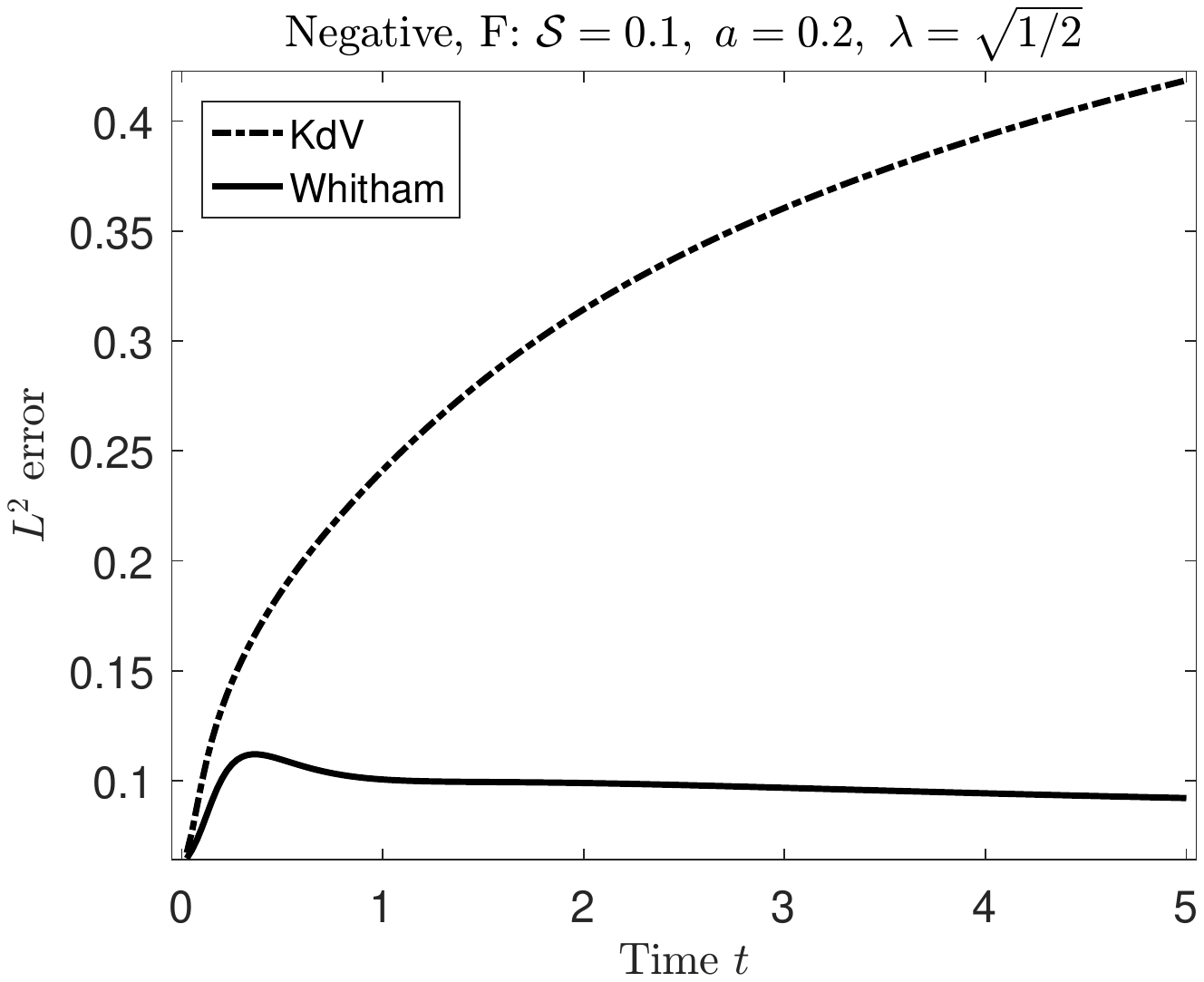}
	}
	\caption
	{
		$L^2$ errors in approximation of solutions
		to the full Euler equations by different model equations
		with the negative initial wave $\eta_0(x)$ and
		the surface tension $\varkappa = \frac 12$.
	}
\label{error_negative_plot_with_tension_frac_12}
\end{figure}
\begin{figure}[t!]
	\centering
	\subfigure
	{
		\includegraphics
		[
			width=0.39\textwidth
			,
			trim = 4.0cm 9cm 4.0cm 9cm
		]
		{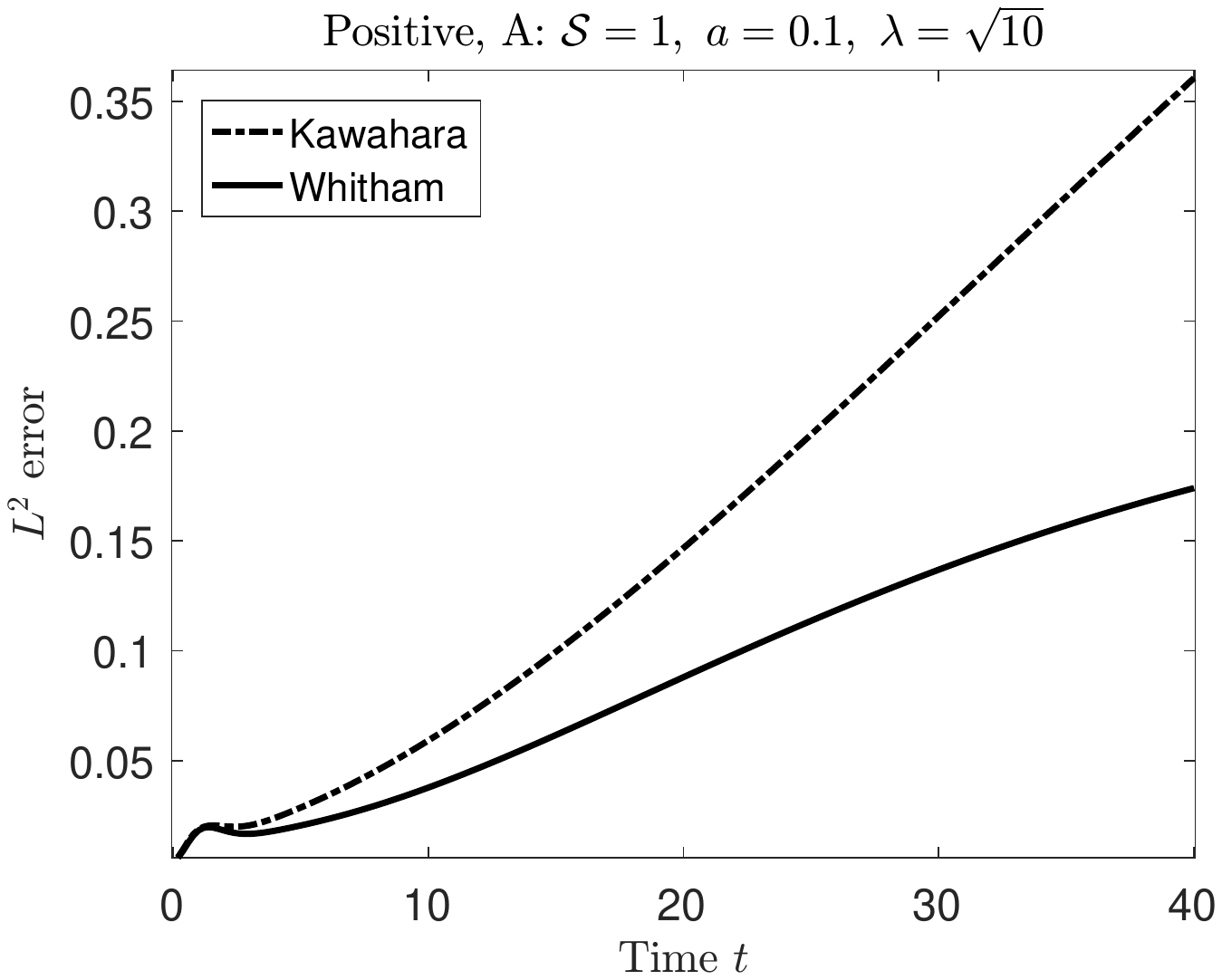}
	}
	~~~~
	\subfigure
	{
		\includegraphics
		[
			width=0.39\textwidth
			,
			trim = 4.0cm 9cm 4.0cm 9cm
		]
		{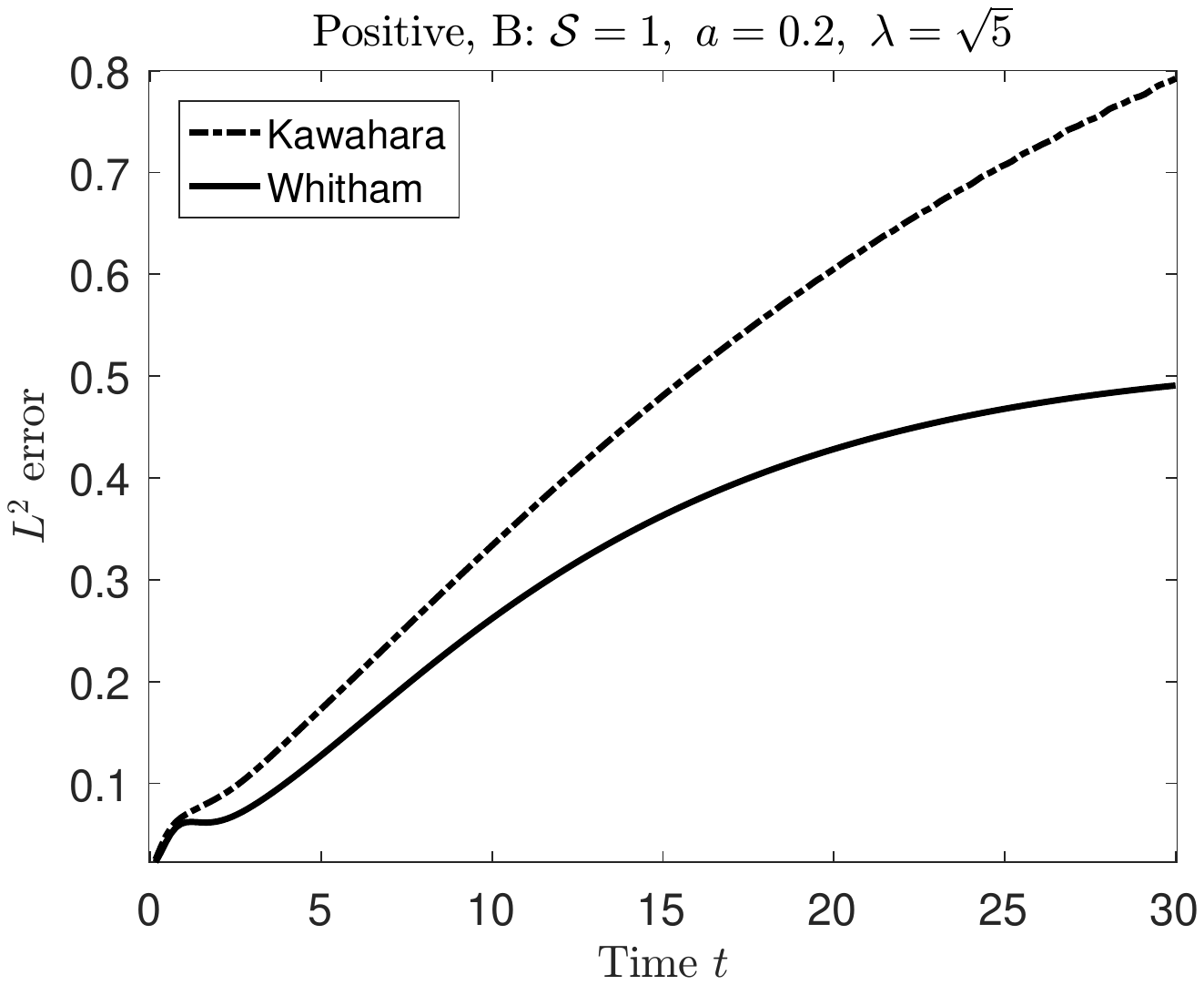}
	}
	\subfigure
	{
		\includegraphics
		[
			width=0.39\textwidth
			,
			trim = 4.0cm 9cm 4.0cm 9cm
		]
		{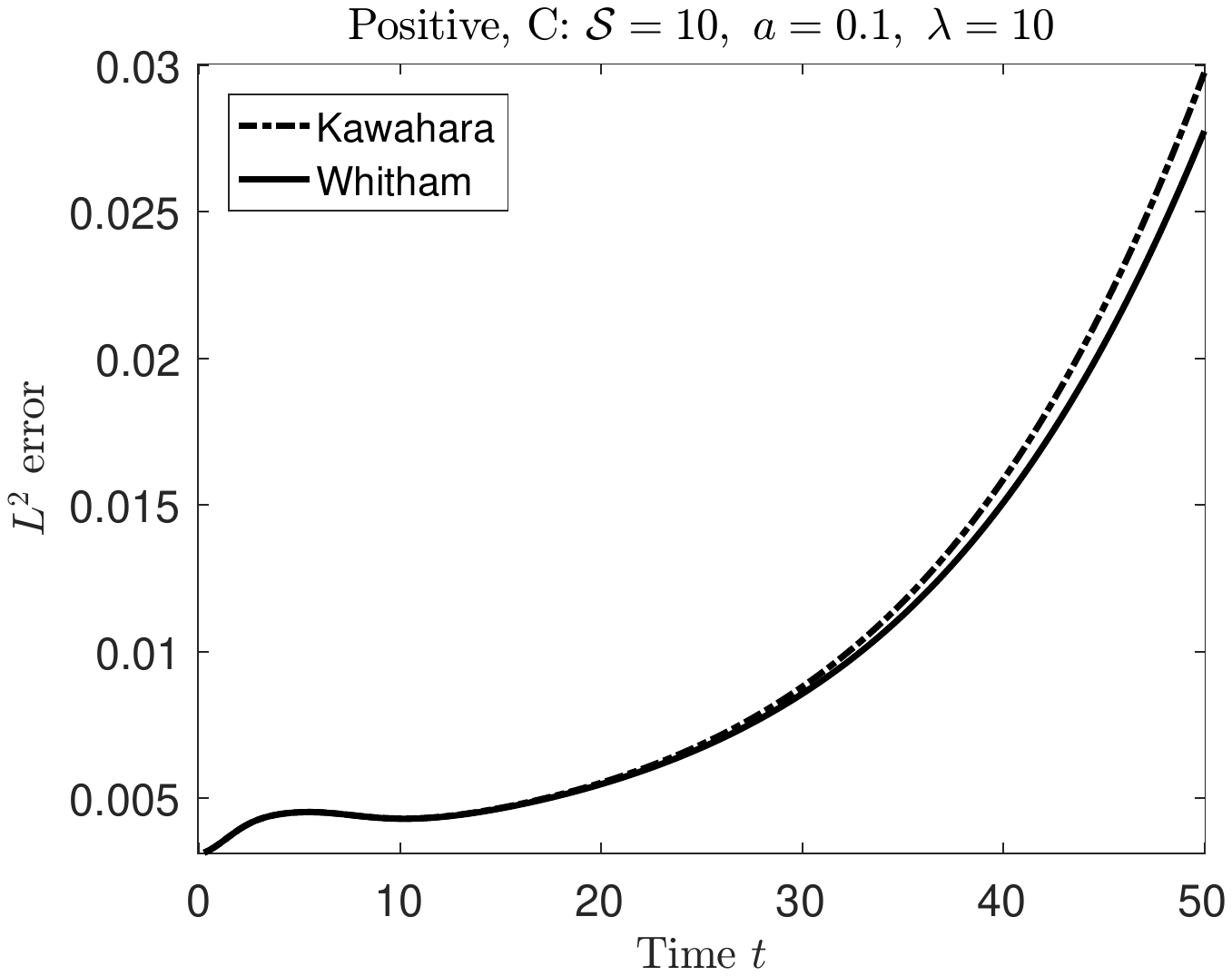}
	}
	~~~~
	\subfigure
	{
		\includegraphics
		[
			width=0.39\textwidth
			,
			trim = 4.0cm 9cm 4.0cm 9cm
		]
		{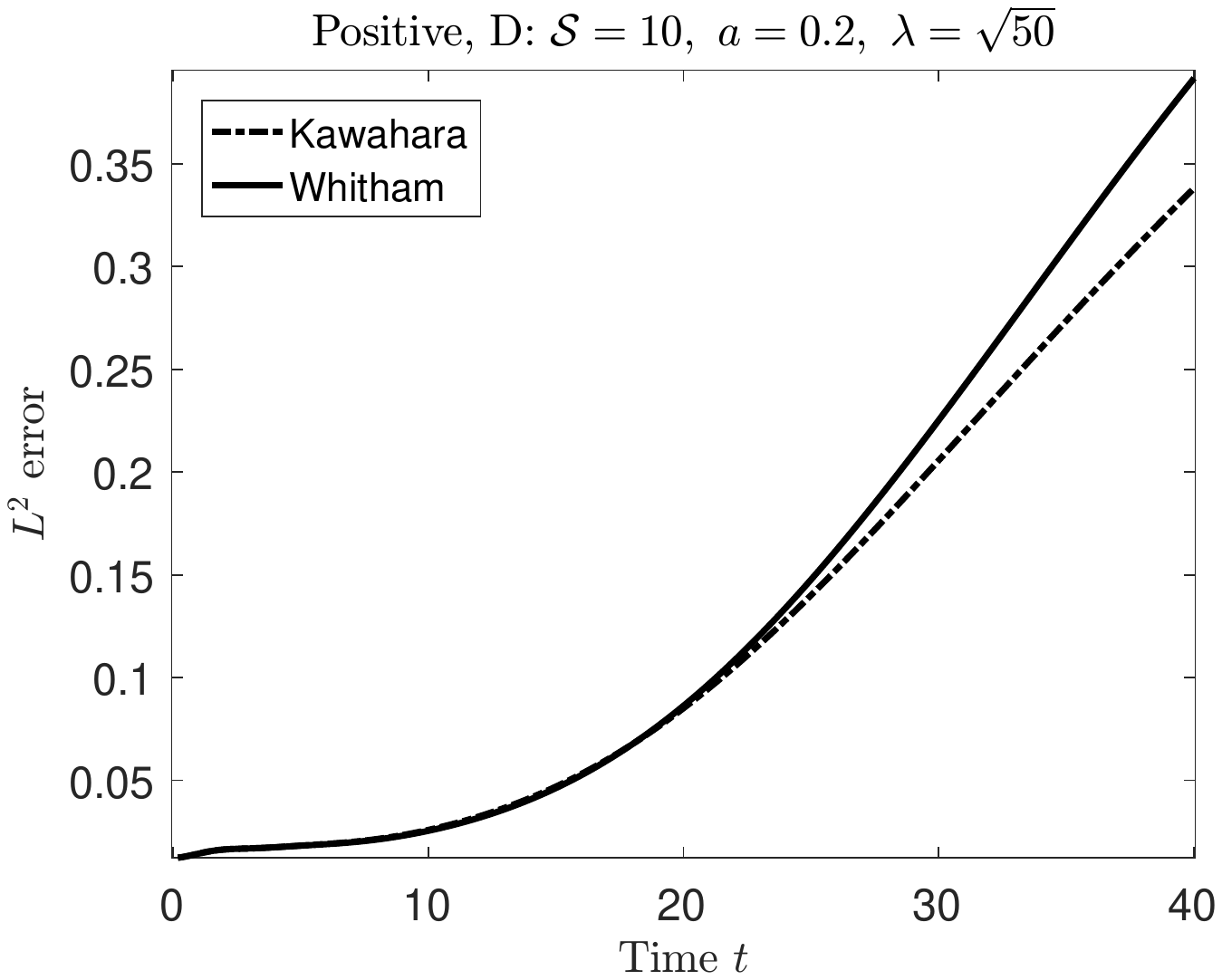}
	}
	\subfigure
	{
		\includegraphics
		[
			width=0.39\textwidth
			,
			trim = 4.0cm 9cm 4.0cm 9cm
		]
		{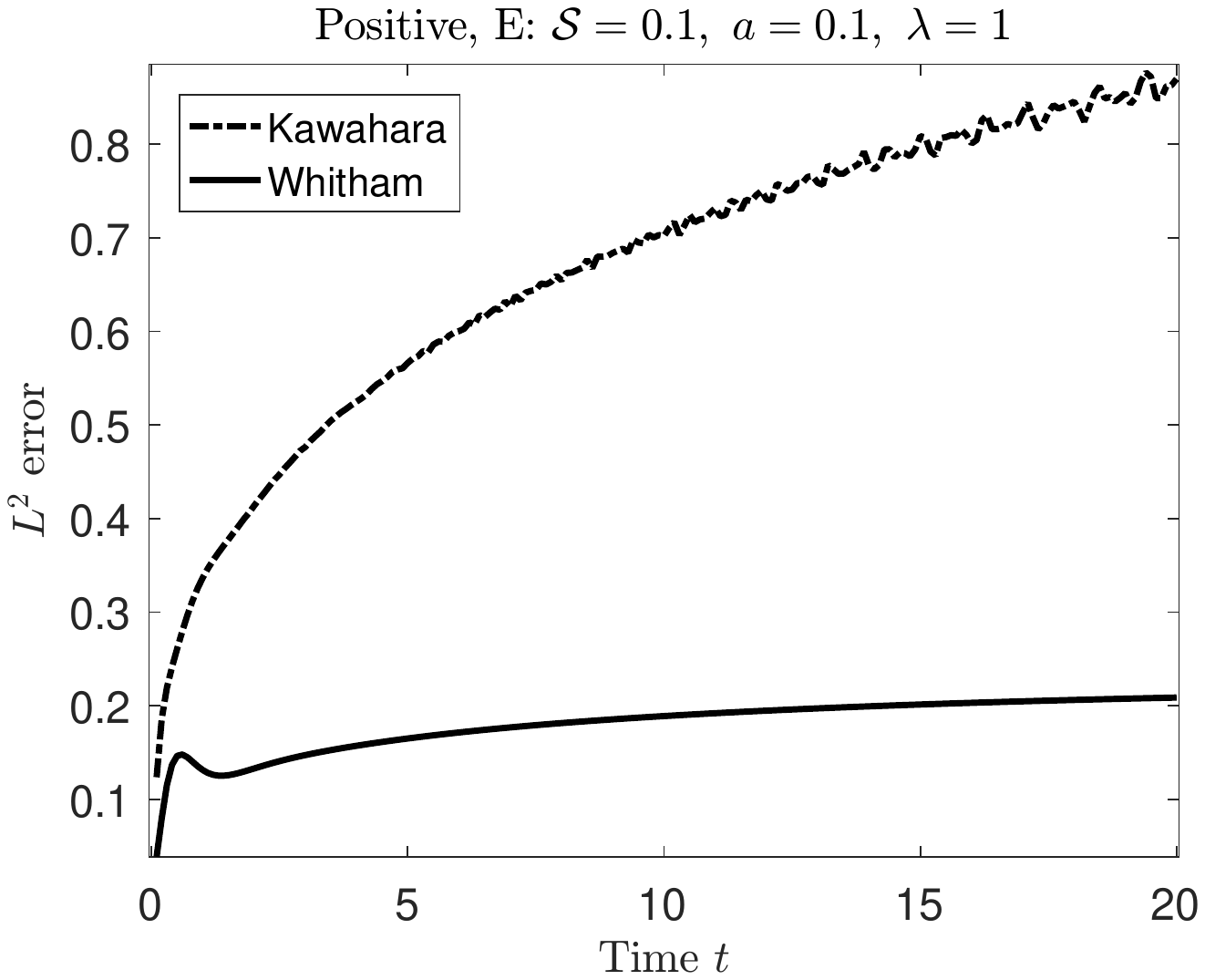}
	}
	~~~~
	\subfigure
	{
		\includegraphics
		[
			width=0.39\textwidth
			,
			trim = 4.0cm 9cm 4.0cm 9cm
		]
		{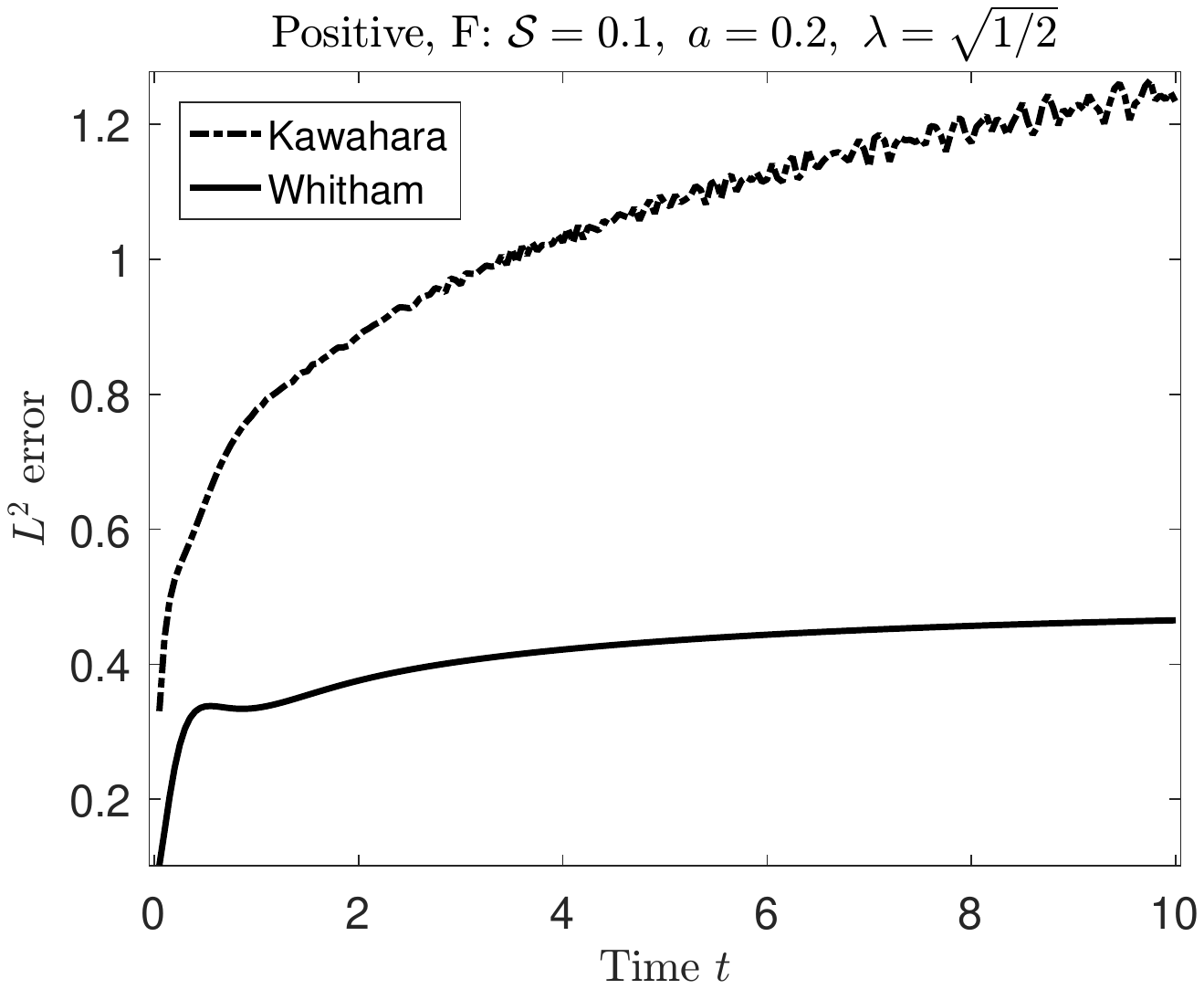}
	}
	\caption
	{
		$L^2$ errors in approximation of solutions
		to the full Euler equations by different model equations
		with the positive initial wave $\eta_0(x)$ and
		the surface tension $\varkappa = \frac 13$.
	}
\label{error_positive_plot_with_tension_frac_13}
\end{figure}
\begin{figure}[t!]
	\centering
	\subfigure
	{
		\includegraphics
		[
			width=0.39\textwidth
			,
			trim = 4.0cm 9cm 4.0cm 9cm
		]
		{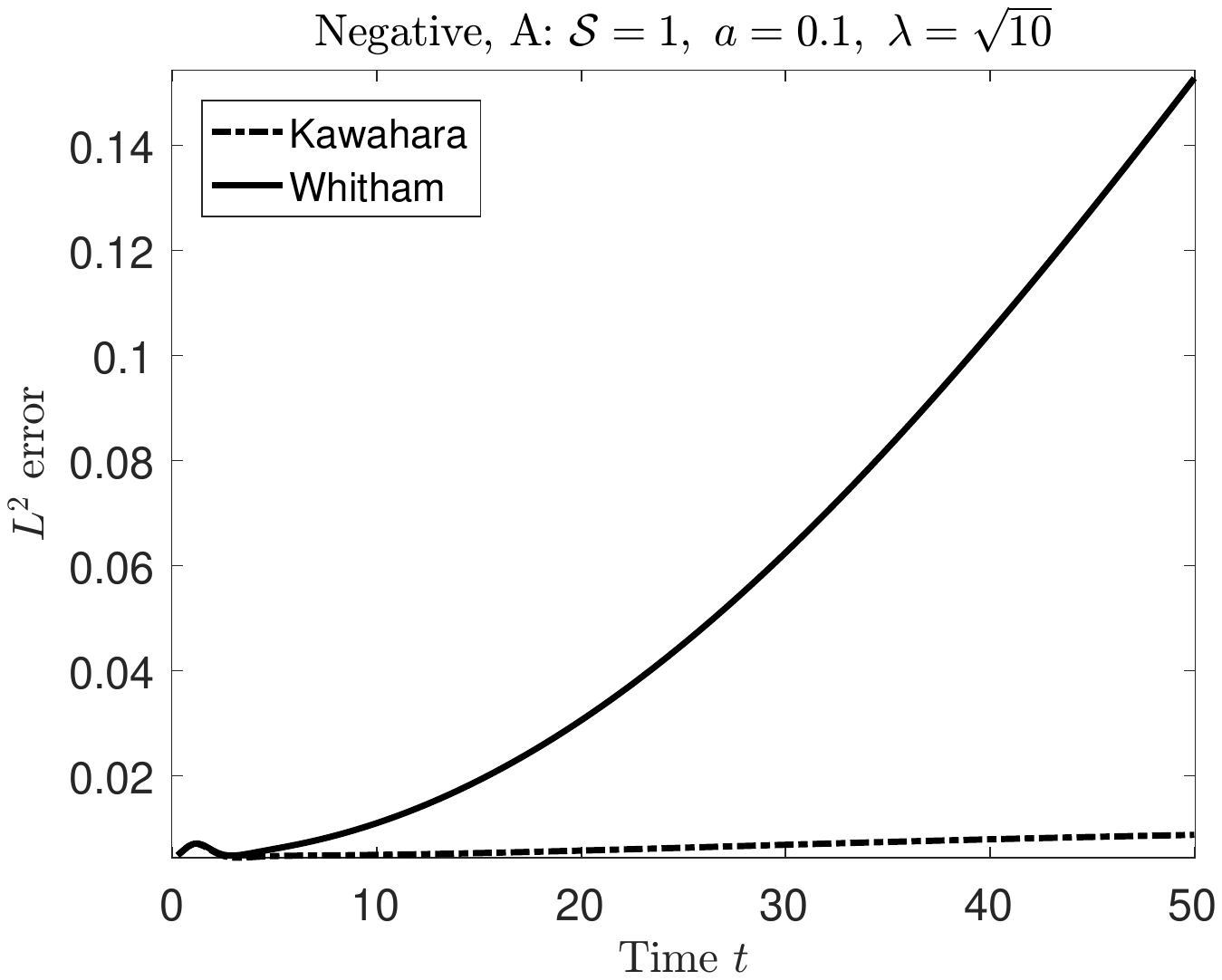}
	}
	~~~~
	\subfigure
	{
		\includegraphics
		[
			width=0.39\textwidth
			,
			trim = 4.0cm 9cm 4.0cm 9cm
		]
		{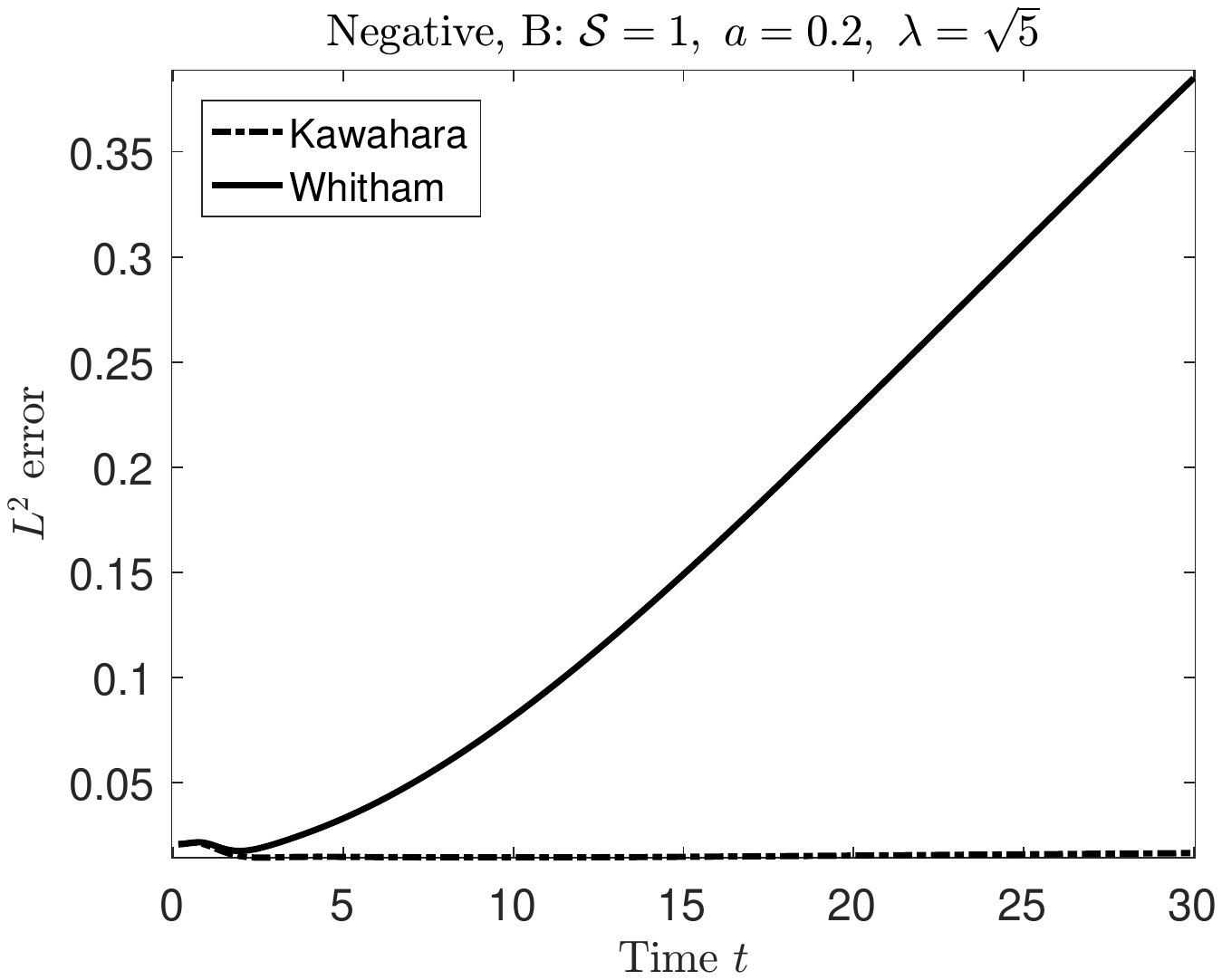}
	}
	\subfigure
	{
		\includegraphics
		[
			width=0.39\textwidth
			,
			trim = 4.0cm 9cm 4.0cm 9cm
		]
		{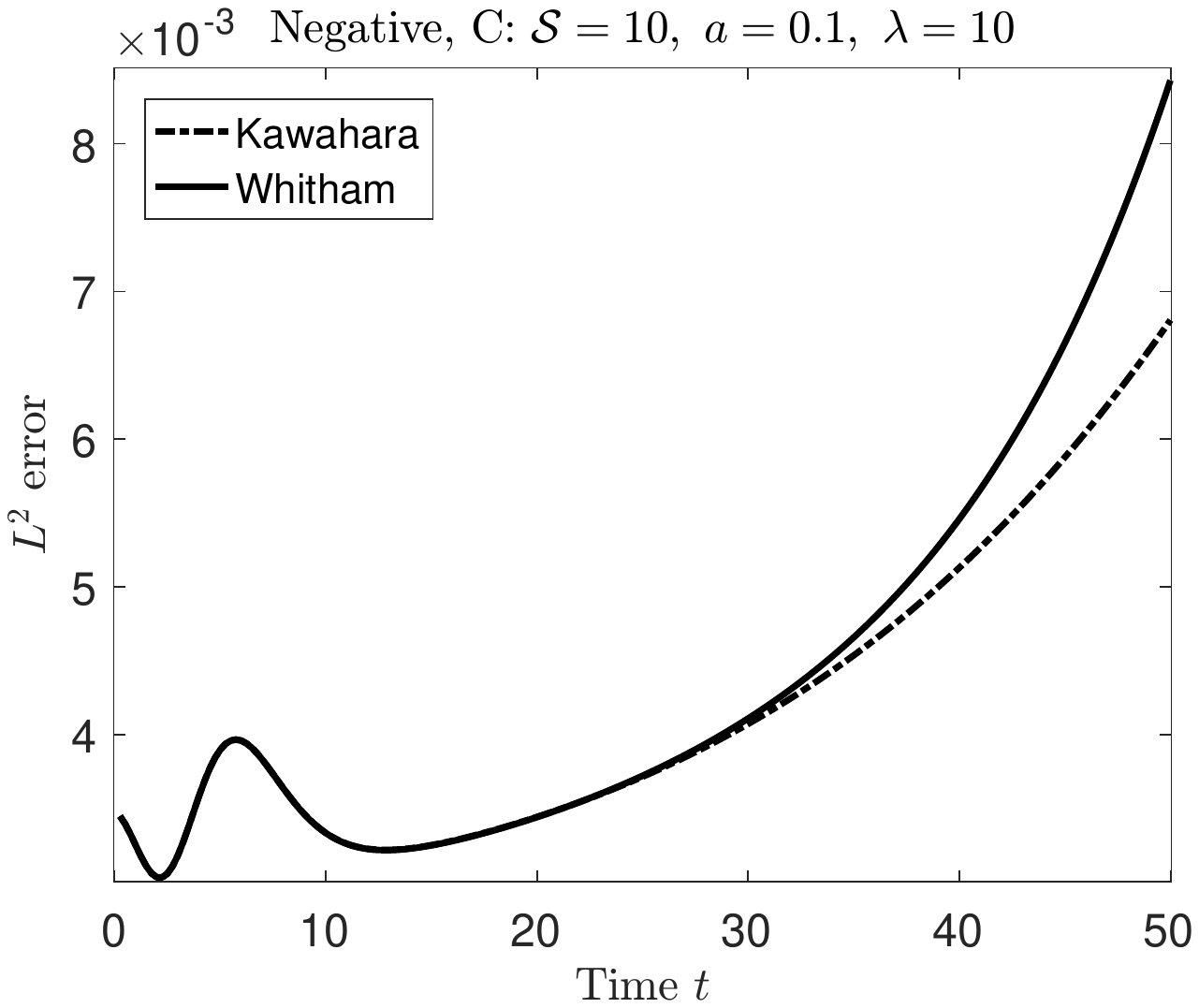}
	}
	~~~~
	\subfigure
	{
		\includegraphics
		[
			width=0.39\textwidth
			,
			trim = 4.0cm 9cm 4.0cm 9cm
		]
		{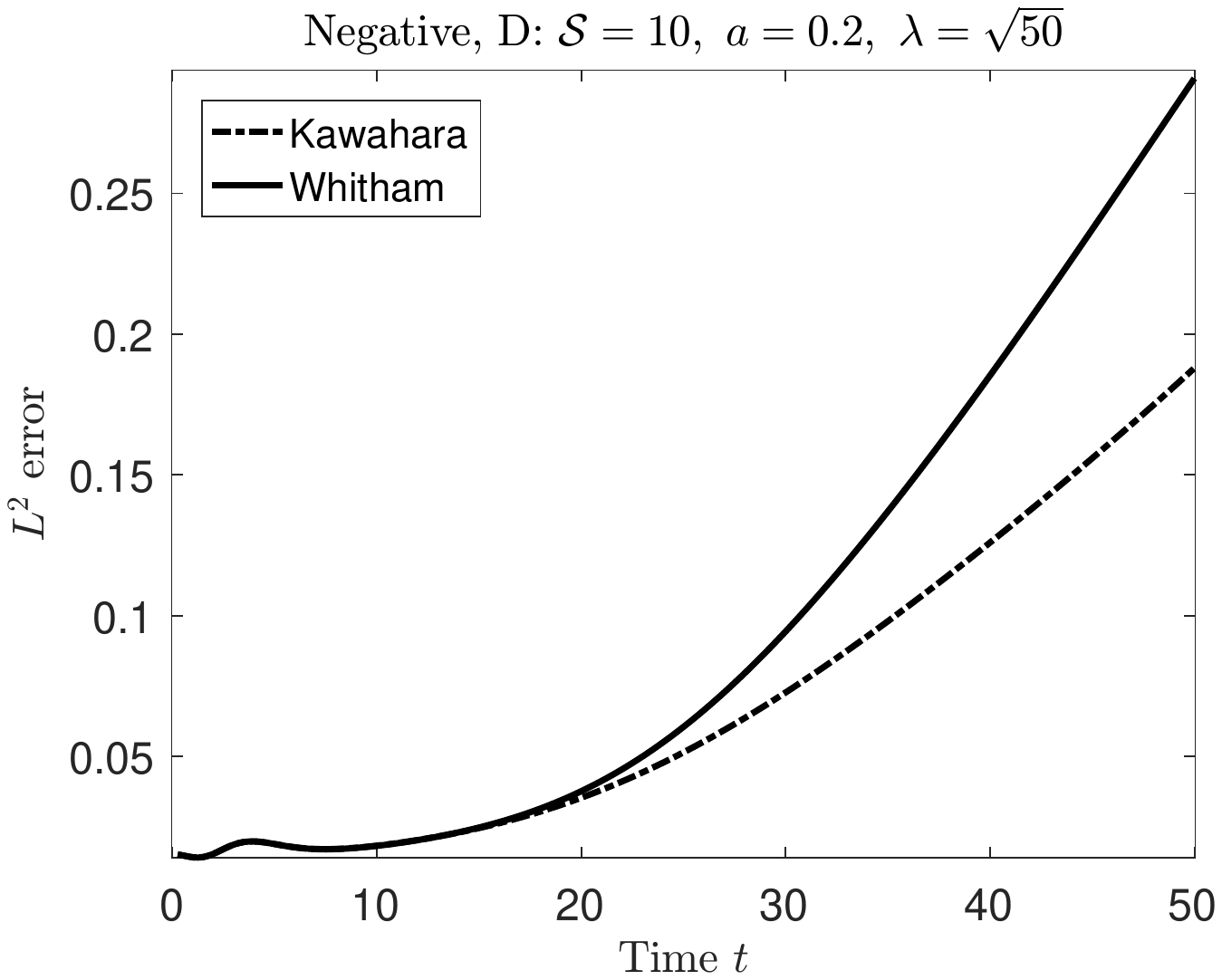}
	}
	\subfigure
	{
		\includegraphics
		[
			width=0.39\textwidth
			,
			trim = 4.0cm 9cm 4.0cm 9cm
		]
		{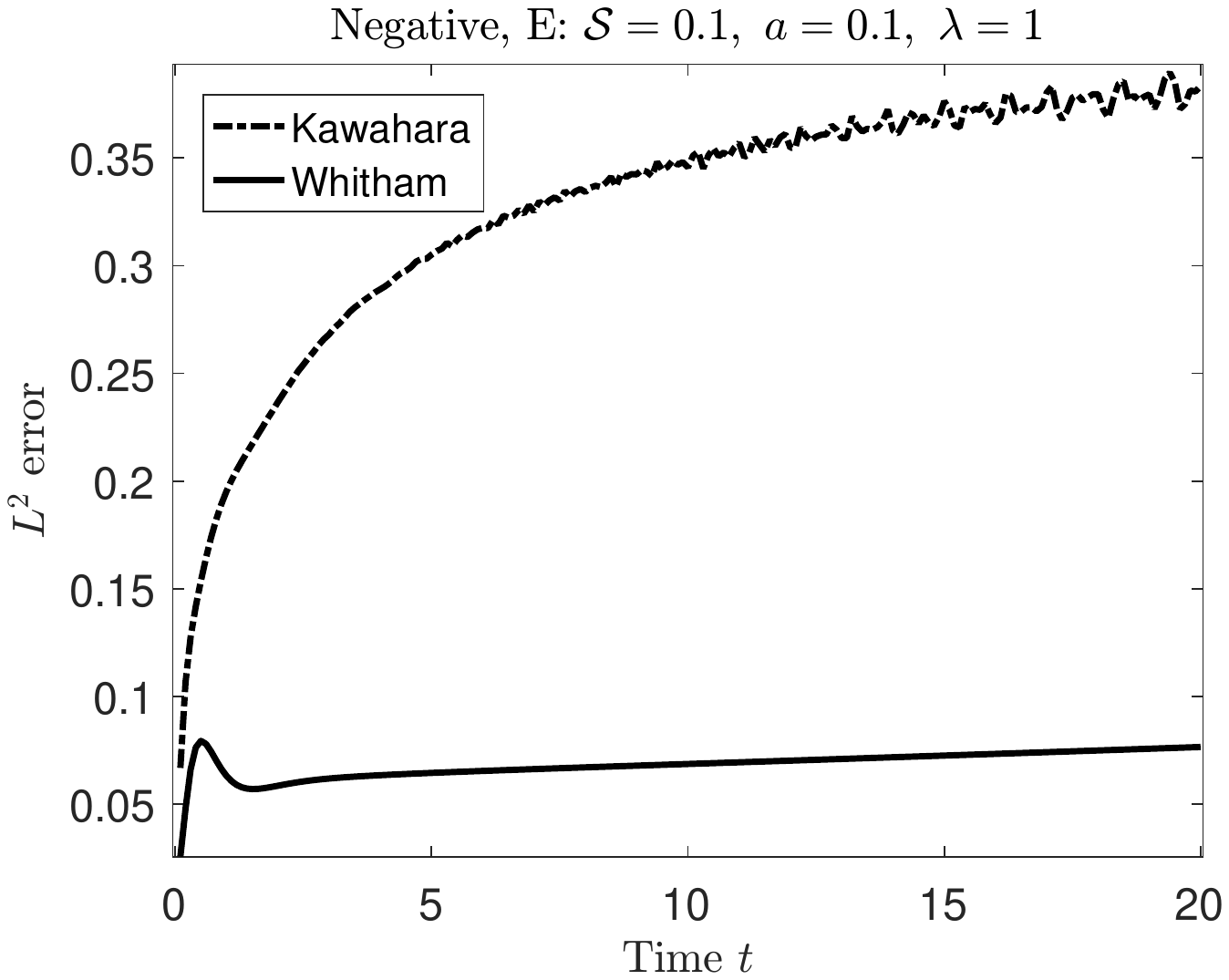}
	}
	~~~~
	\subfigure
	{
		\includegraphics
		[
			width=0.39\textwidth
			,
			trim = 4.0cm 9cm 4.0cm 9cm
		]
		{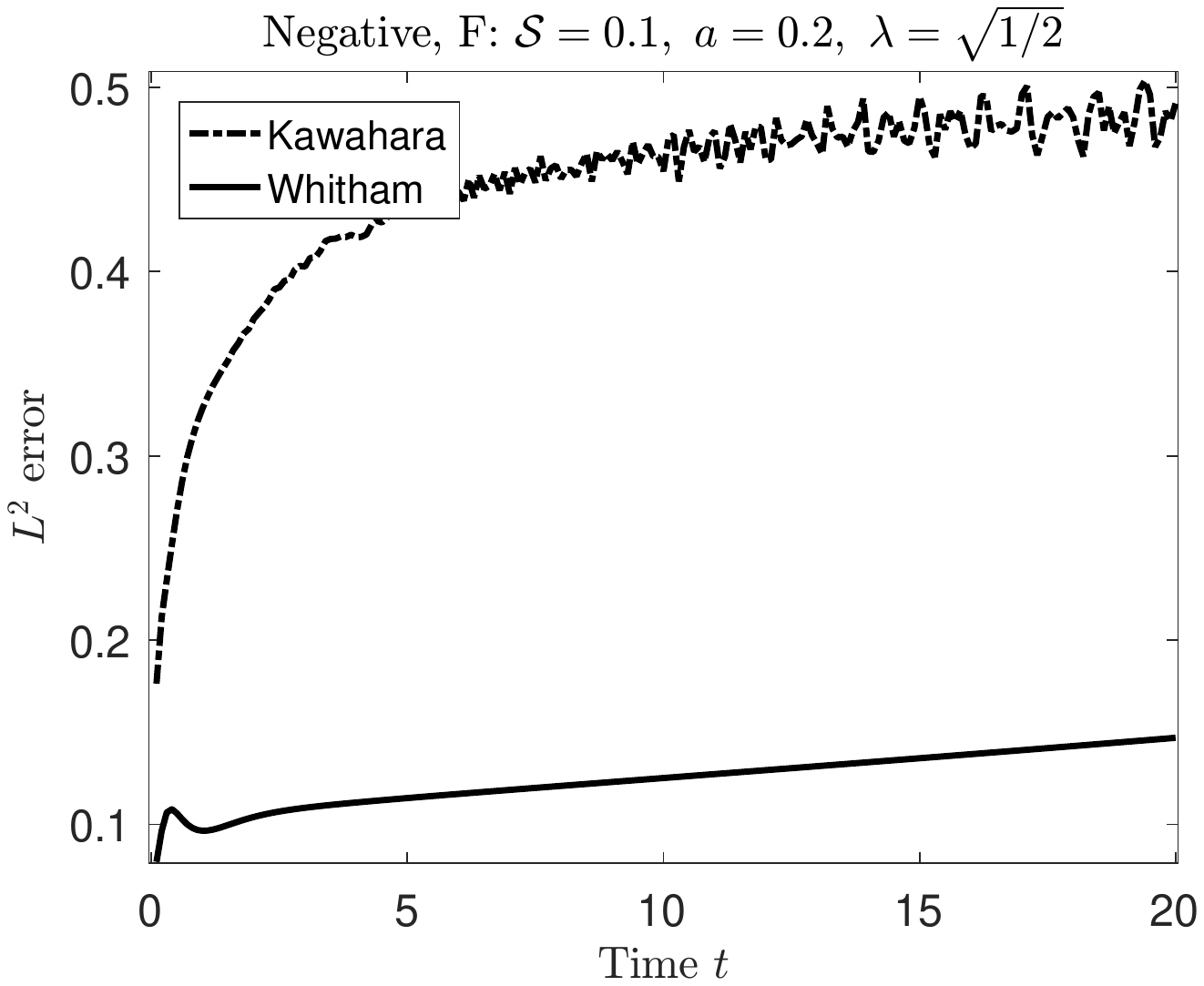}
	}
	\caption
	{
		$L^2$ errors in approximation of solutions
		to the full Euler equations by different model equations
		with the negative initial wave $\eta_0(x)$ and
		the surface tension $\varkappa = \frac 13$.
	}
\label{error_negative_plot_with_tension_frac_13}
\end{figure}

Initial conditions for the Euler equations are chosen in such a way that 
the solutions are expected to be right moving.
This can be achieved by
imposing an initial surface disturbance $\eta_0$ together with the initial
trace of the potential
\(
	\Phi(x) = \int_0^x \eta_0(\xi) \, d\xi
	.
\)
Indeed, the right-going wave condition is $s(x,t) = 0$ which together with
\eqref{variable_transformation} and \eqref{K_transformation} imply
\[
	\Phi(x) = \int_0^x u(\xi, 0) d\xi
	= \int_0^x K^{-1} \eta_0 (\xi) d\xi
	\approx \int_0^x \eta_0 (\xi) d\xi
	.
\]
This last provision makes our numerical experiments more natural,
since it is not assumed that the regarded surface waves are
strictly right-moving $s = O(\mu^2 \alpha)$ and $\eta = r + O(\mu^2 \alpha)$.
In Figure \ref{motion_plot} one can see the corresponding small
wave $s$ moving to the left given by solving the Euler system.
In order to normalize the data, we choose $\eta_0$ in such a way that
the average of $\eta_0$ over the computational domain is zero.
The experiments are performed with
several different amplitudes $\alpha$ and wavelengths $\lambda$.
For the purpose of this section, we define the 
wavelength $\lambda$ as the distance between 
the two points $x_1$ and $x_2$ at which $\eta_0(x_1) = \eta_0(x_2) = \alpha /2$.
Both positive and negative initial disturbances are considered.
Numerical experiments were performed with a range of parameters
for amplitude $\alpha$ and the wave-length $\lambda$.
The summary of experiments' settings is given in Table 1.
All experiments are made with initial wave of elevation and wave of depression, labeled as ``positive" and ``negative" respectively.
The domain for computations is $-L\leq x \leq L$, with $L = 100$. 
The ``positive" initial data is
\begin{equation}
\label{test_1}
	\eta_0(x) = a\cdot \sech^2(f(\lambda)x)-C, 
\end{equation}
where 
\[
	f(\lambda) = \frac{2}{ \lambda }
	\log \left( \frac{ 1 + \sqrt{1/2} }{ \sqrt{1/2} } \right)
	, \qquad
	C = \frac{1}{L}\frac{a}{f(\lambda)}\tanh \left(\frac{L}{f(\lambda)}\right)
	.
\]
Here $C$ and $f(\lambda)$ are chosen so that 
$\int_{-L}^{L}\eta_0(x)dx = 0$, and
the wave-length $\lambda$
is the distance between the two points $x_1$ and $x_2$
at which $\eta_0(x_1) = \eta_0(x_2) = a/2$.
The velocity potential in this case is:
\begin{equation}
\label{velocity1}
	\Phi(x) = \frac{a}{f(\lambda)}\tanh(f(\lambda)x)-Cx
	,
\end{equation}
The ``negative" case function is just the ``reverse" of the first one
\begin{equation}
\label{test_2}
	\eta_0(x) = - a\cdot \sech^2(f(\lambda)x) + C
	. 
\end{equation}
The definitions for $f(\lambda)$ and $C$ are the same.
And the velocity potential is
\begin{equation}
\label{velocity2}
	\Phi(x) = -\frac{a}{f(\lambda)}\tanh(f(\lambda)x)+Cx
	,
\end{equation}
We calculate solutions of the Whitham equation and the Euler system.
We also calculate solutions of the KdV for the capillarity $\varkappa = 1/2$,
and solutions of the Kawahara in the case $\varkappa = 1/3$.

In Figure \ref{motion_plot},
the time evolution of a wave with an initial narrow peak is shown according to
the Euler (black line), Whitham (red line) and KdV (blue line) equations. 
Here the amplitude $\alpha = 0.2$,
the wavelength $\lambda = \sqrt{5}$ and the capillary parameter $\varkappa = 1/2$
are used.
This case corresponds the Stokes number $\mathcal S = 1$.
It appears that the KdV equation
produces a significant amount of spurious oscillations
and the Whitham equation gives the closest approximation of
the corresponding Euler solution.
As one can expect the unidirectional models 
also lag in the description of
waves going to the left.

In order to compare the accuracy of each approximate model
we calculate the differences, firstly between the Whitham and Euler equations,
and secondly between the KdV (or Kawahara) and Euler equations.
These differences are measured in the integral $L^2$-norm
normalized by initial condition $\lVert \eta_0 \rVert$ as follows
\[
	\frac{ \lVert \eta_E(t) - \eta(t) \rVert }{ \lVert \eta_0 \rVert }
	=
	\sqrt
	{
		\frac{ \int ( \eta_E(x, t) - \eta(x, t) )^2 dx }{ \int \eta_0(x)^2 dx }
	}
\]
where $ \eta_E(x, t) $ is the solution for the Euler system
and $ \eta(x, t) $ corresponds either to the Whitham, KdV or Kawahara equation.
The next figures \ref{error_positive_plot_with_tension_frac_12},
\ref{error_negative_plot_with_tension_frac_12},
\ref{error_positive_plot_with_tension_frac_13},
\ref{error_negative_plot_with_tension_frac_13}
show the dependence of $L^2$-error on time for different
initial situations.
Thus as one can see, the Whitham model
performs better then the KdV and Kawahara equations,
in nearly all situations except the cases with a negative initial wave of depression
and Stokes number approximately unity.

\vskip 0.05in
\noindent
{\bf Acknowledgments.}

This research was supported in part by the Research Council of Norway under grant no. 213474/F20 and grant no. 239033/F20.


\vskip -0.1in
\begin{thebibliography}{26}
\small \setlength{\itemsep}{-0.4mm}
%
%

\bibitem{AMP}
Aceves-S\'{a}nchez, P., Minzoni, A.A. and Panayotaros, P. 
{\em Numerical study of a nonlocal model for water-waves with variable depth}, 
Wave Motion {\bf 50} (2013), 80--93.


\bibitem{Alazard}
Alazard, T., Burq, N. and Zuily, C.
{\em On the water-wave equations with surface tension}, 
Duke Math. Journal {\bf 158} (2011), 413--99.


\bibitem{Biswas} 
Biswas A.
{\em Solitary wave solution for the generalized Kawahara equation}, 
Appl. Math. Lett. {\bf 22} (2009), 208--210. 

\bibitem{BBM}
Benjamin, T. B., Bona, J. L. and Mahony, J. J. 
{\em Model equations for long waves in nonlinear dispersive systems}. 
Philos. Trans. R. Soc. Lond., Ser. A {\bf 272} (1972), 47--78.


\bibitem{BKN} Borluk, H., Kalisch, H. and Nicholls, D.P. 
{\em A numerical study of the Whitham equation as a model for steady surface water waves}, 
J. Comput. Appl. Math. {\bf 296} (2016) 293--302. 

\bibitem{Chardard} Chardard, F. {\em Stabilit\'{e} des ondes solitaires}, PhD thesis, 2009.

\bibitem{CC1999}
Choi, W. and Camassa, R. 
{\em Exact Evolution Equations for Surface Waves}. 
J. Eng. Mech. {\bf 125} (1999), 756--760.

\bibitem{CG} 
Craig, W. and Groves, M.D. 
{\em Hamiltonian long-wave approximations to the water-wave problem}. 
Wave Motion {\bf 19} (1994), 367--389. 

\bibitem{CGK} 
Craig, W., Guyenne, P. and Kalisch, H. 
{\em Hamiltonian long-wave expansions for free surfaces and interfaces}.
Comm. Pure Appl. Math. {\bf 58} (2005), 1587--1641.


\bibitem{CS}
Craig, W. and Sulem, C. 
{\em Numerical simulation of gravity waves}.
J. Comp. Phys. {\bf 108} (1993), 73--83.

\bibitem{DyachenkoZakharov}
Dyachenko, A.I., Kuznetsov, E.A., Spector, M.D. and  Zakharov, V.E.
{\em Analytical description of the free surface dynamics of an ideal fluid (canonical formalism and conformal mapping)}. 
Phys. Lett. A {\bf 221} (1996), 73--79.

\bibitem{FruSan} 
De Frutos, J. and Sanz-Serna, J.M. 
{\em An easily implementable fourth-order method for the time integration of wave problems}.
J. Comp. Phys. {\bf 103} (1992), 160--168.


\bibitem{EGW}
Ehrnstr{\"o}m, M., Groves, M.D. and  Wahl\'en, E. 
{\em Solitary waves of the Whitham equation - a variational approach to a class of nonlocal
  evolution equations and existence of solitary waves of the Whitham equation}.
Nonlinearity {\bf  25} (2012 ), 2903--2936.

\bibitem{EK1}
Ehrnstr{\"o}m, M. and Kalisch, H. 
{\em Traveling waves for the Whitham equation}. 
Differential Integral Equations {\bf 22} (2009), 1193--1210.

\bibitem{EK2}
Ehrnstr{\"o}m, M. and Kalisch, H. 
{\em Global bifurcation for the Whitham equation}.
Math. Mod. Nat. Phenomena {\bf 8} (2013), 13--30.


\bibitem{FW}
Fornberg, B. and Whitham, G.B. 
{\em A Numerical and Theoretical Study of Certain Nonlinear Wave Phenomena}. 
Phil. Trans. Roy. Soc. A {\bf 289} (1978 ), 373--404.


\bibitem{Hammack}
Hammack, J.L. and Segur, H. 
{\em The Korteweg-de Vries equation and water waves.
Part 2. Comparison with experiments}.
J. Fluid Mech. {\bf 65} (1974), 289-314.


\bibitem{HJ1}
Hur, V.M. and Johnson, M. 
{\em Modulational instability in the Whitham equation of water waves}.
Studies in Applied Mathematics {\bf 134} (2015), 120--143.

\bibitem{HJ2}
Hur, V.M. and Johnson, M. 
{\em Modulational instability in the Whitham equation with surface tension and vorticity}.
Nonlinear Anal. {\bf 129} (2015), 104--118.


\bibitem{Kawahara} Kawahara, T. 
{\em Oscillatory solitary waves in dispersive media}. 
J. Phys. Soc. Japan {\bf 33} (1972), 260--264.


\bibitem{KoopButler}
Koop, C.G. and Butler, G. 
{\em An investigation of internal solitary waves in a two-fluid system}. 
J. Fluid Mech., {\bf 112} (1981), 225--251.

\bibitem{LannesBOOK} 
Lannes, D. 
The Water Waves Problem.
Mathematical Surveys and Monographs, vol. \textbf{188} 
(Amer. Math. Soc., Providence, 2013).


\bibitem{LS} Lannes, D. and Saut, J.-C.
{\em Remarks on the full dispersion Kadomtsev-Petviashvli equation},
Kinet. Relat. Models {\bf 6} (2013), 989--1009.


\bibitem{LHCh}
Li, Y.A., Hyman, J.M. and Choi, W.
{\em A Numerical Study of the Exact Evolution Equations for Surface Waves in Water of Finite Depth}. 
Stud. Appl. Math. {\bf 113} (2004), 303--324.

\bibitem{Milewski}
Milewski, P., Vanden-Broeck, J.-M. and Wang, Z. 
{\em Dynamics of steep two-dimensional gravity-capillary solitary waves}. 
J. Fluid Mech. {\bf 664} (2010), 466--477.

\bibitem{MDC}
Mitsotakis, D., Dutykh, D. and Carter, J.D. 
{\em On the nonlinear dynamics of the traveling-wave solutions of the Serre equations}. 
Wave Motion, to appear.

\bibitem{MKD} Moldabayev, D., Kalisch, H. and  Dutykh, D. 
{\em The Whitham Equation as a model for surface water waves},
Phys. D  {\bf 309} (2015), 99--107.

\bibitem{Nicholls}
Nicholls, D.P. and  Reitich, F. 
{\em A new approach to analyticity of Dirichlet-Neumann operators}.
Proc. Roy. Soc. Edinburgh Sect. A {\bf 131} (2001), 1411--1433. 

\bibitem{Ovsyannikov}
Ovsyannikov, L.V. 
{\em To the shallow water theory foundation}. 
Arch. Mech. {\bf 26} (1974), 407--422.

\bibitem{Pe} 
Peregrine, D.H. 
{\em Calculations of the development of an undular bore}. 
J. Fluid Mech. {\bf 25} (1966), 321--330.

\bibitem{Petrov}
Petrov, A.A. 
{\em Variational statement of the problem of liquid motion in a container of finite dimensions}. 
Prikl. Math. Mekh. {\bf 28} ( 1964 ), 917--922.

\bibitem{Sanford}
Sanford, N., Kodama, K., Carter, J. D. and Kalisch, H. 
{\em Stability of traveling wave solutions to the Whitham equation}. 
Phys Lett. A {\bf 378} (2014), 2100--2107.

\bibitem{Stepanyants} Stepanyants, Y.
{\em Dispersion of long gravity-capillary surface waves and asymptotic equations for solitons}, 
Proceedings of the Russian Academy of Engineering Sciences Series: Applied Mathematics and Mechanics
{\bf 14} (2005), 33-40.


\bibitem{Wh1} 
Whitham, G.B. 
{\em Variational methods and applications to water waves}.
Proc. Roy. Soc. London A {\bf 299} (1967), 6--25.


\bibitem{Wh2}
Whitham, G.B. 
Linear and Nonlinear Waves (Wiley, New York, 1974).


\bibitem{Zh} 
Zakharov, V.E. 
{\em Stability of periodic waves of finite amplitude on the surface of a deep fluid}. 
J. Appl. Mech. Tech. Phys. {\bf 9} (1968), 190--194.


\end{thebibliography}
\end{document}